\documentclass[USenglish,oneside,twocolumn]{article}

\mathchardef\Gamma="0100 \mathchardef\Delta="0101
\mathchardef\Theta="0102 \mathchardef\Lambda="0103
\mathchardef\Xi="0104 \mathchardef\Pi="0105
\mathchardef\Sigma="0106 \mathchardef\Upsilon="0107
\mathchardef\Phi="0108 \mathchardef\Psi="0109
\mathchardef\Omega="010A

\newcommand{\outline}[1]{}%{\textbf{#1}}

\usepackage{cite}
\usepackage{xspace}
\usepackage{graphicx}
\usepackage{latexsym}
\usepackage{amssymb}
\usepackage{amsfonts}
\usepackage{psfrag}
\usepackage{wrapfig}
\usepackage{comment}
\usepackage{alltt}
\usepackage{color}

\newcommand{\Comment}[1]{}

\usepackage{pifont}% http://ctan.org/pkg/pifont

\usepackage[utf8]{inputenc}%(only for the pdftex engine)
\usepackage[big]{dgruyter_NEW}
\usepackage{algorithm}
\usepackage{algpseudocode}
\usepackage{enumitem,epstopdf}
\usepackage{bbm}
\usepackage{balance}
\usepackage{graphicx}
\usepackage{wrapfig}
\usepackage{longtable}
\usepackage{epstopdf}
\usepackage{soul,url}
\usepackage{amsmath}
\usepackage{float}
\usepackage{comment,caption}
\usepackage{booktabs,balance}
\usepackage{tabu,multirow}
\usepackage{blindtext}
\usepackage{hyperref}
\usepackage{graphicx}
\usepackage{caption}
\usepackage{subcaption}
\usepackage[dvipsnames]{xcolor}

\usepackage{slashbox}
\usepackage{mathrsfs}
\usepackage[framemethod=TikZ]{mdframed}
\usepackage{listings}
\usepackage{xcolor}

\makeatletter
\def\BState{\State\hskip-\ALG@thistlm}
\makeatother

\cclogo{\includegraphics{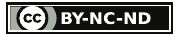}}

\begin{document}

  \author[1]{Yuantian Miao}

  \author[2]{Minhui Xue}

  \author*[3]{Chao Chen}

  \author[4]{Lei Pan}

  \author[5]{Jun Zhang}
  
  \author[6]{Benjamin Zi Hao Zhao}

  \author[7]{Dali Kaafar}

  \author[8]{Yang Xiang}

  \affil[1]{Swinburne University of Technology, Australia}

  \affil[2]{The University of Adelaide, Australia}

  \affil[3]{James Cook University, Australia. Email: chao.chen@jcu.edu.au.}

  \affil[4]{Deakin University, Australia}

  \affil[5]{Swinburne University of Technology, Australia}

  \affil[6]{The University of New South Wales and CSIRO-Data61, Australia}

  \affil[7]{Macquarie University and CSIRO-Data61, Australia}

  \affil[8]{Swinburne University of Technology, Australia}

\title{\huge The Audio Auditor: User-Level Membership Inference in Internet of Things Voice Services}
\runningtitle{The Audio Auditor: User-Level Membership Inference in Internet of Things Voice Services}

\begin{abstract}
{With the rapid development of deep learning techniques, the popularity of voice services implemented on various Internet of Things (IoT) devices is ever increasing. In this paper, we examine user-level membership inference in the problem space of voice services, by designing an audio auditor to verify whether a specific user had unwillingly contributed audio used to train an automatic speech recognition (ASR) model under strict black-box access. With user representation of the input audio data and their corresponding translated text, our trained auditor is effective in user-level audit. We also observe that the auditor trained on specific data can be generalized well regardless of the ASR model architecture. We validate the auditor on ASR models trained with LSTM, RNNs, and GRU algorithms on two state-of-the-art pipelines, the hybrid ASR system and the end-to-end ASR system. Finally, we conduct a real-world trial of our auditor on iPhone Siri, achieving an overall accuracy exceeding 80\%. We hope the methodology developed in this paper and findings can inform privacy advocates to overhaul IoT privacy.}
\end{abstract}
\keywords{Membership Inference Attack, ASR, Machine Learning}

\journalname{Proceedings on Privacy Enhancing Technologies}
% \DOI{Editor to enter DOI}
% \startpage{1}
% \received{..}
% \revised{..}
% \accepted{..}

\journalyear{2021}

\maketitle

\section{Introduction} \label{sec:intro}

Automatic speech recognition (ASR) systems are widely adopted on Internet of Things (IoT) devices~\cite{lokesh2018automatic, mehrabani2015personalized}. In the IoT voice services space, competition in the smart speaker market is heating up between giants like Apple, Microsoft, and Amazon~\cite{amazon2017verge}. However parallel to the release of new products, consumers are growing increasingly aware and concerned about their privacy, particularly about unauthorized access to user's audio in these ASR systems. Of late, privacy policies and regulations, such as the General Data Protection Regulations (GDPR)~\cite{GDPR}, the Children's Online Privacy Protection Act (COPPA)~\cite{mcreynolds2017toys}, and the California Consumer Privacy Act (CCPA)~\cite{ccpa2020california}, have been enforced to regulate personal data processing. Specifically, the Right to be Forgotten~\cite{xue2016right} law allows customers to prevent third-party voice services from continuously using their data~\cite{BBC2019voice}. However, the murky boundary between privacy and security can thwart IoT's trustworthiness~\cite{how2019news,voice2019news} and many IoT devices may attempt to sniff and analyze the audio captured in real-time without a user's consent~\cite{alexa2018}. Most recently, on WeChat -- a hugely popular messaging platform within China and Worldwide -- a scammer camouflaged their voice to sound like an acquaintance by spoofing his or her voice~\cite{wechat2018}. Additionally, in 2019, The Guardian reported a threat regarding the user recordings leakage via Apple Siri~\cite{guardian2019apple}. Auditing if an ASR service provider adheres to its privacy statement can help users to protect their data privacy. It motivates us to develop techniques that enable auditing the use of customers' audio data in ASR models.

Recently, researchers have shown that \textit{record-level membership inference}~\cite{shokri2017membership,hayes2019logan,salem2019ml} may expose information about the model's training data even with only black-box access. To mount membership inference attacks, Shokri et al.~\cite{shokri2017membership} integrate a plethora of shadow models to constitute the attack model to infer membership, while Salem et al.~\cite{salem2019ml} further relax this process and resort to the target model's confidence scores alone. However, instead of inferring record-level information, we seek to infer user-level information to verify whether a user has any audios within the training set. Therefore, we define \textbf{user-level membership inference} as: \textit{querying with a user's data, if this user has any data within target model's training set, even if the query data are not members of the training set, this user is the user-level member of this training set.}

Song and Shmatikov~\cite{song2019auditing} discuss the application of user-level membership inference on text generative models, exploiting several top ranked outputs of the model. 
Considering most ASR systems in the real world do not provide the confidence score, significantly differing from text generative models lending confidence scores~\cite{song2019auditing}, this paper targets user-level membership inference on ASR systems under \textit{strict black-box access}, which we define as no knowledge about the model, with only knowledge of the model's output \textit{excluding confidence score and rank information}, i.e., only predicted label is known.

%%%% Challenge in user-level Membership  
Unfortunately, user-level membership inference on ASR systems with strict black-box access is challenging. (i) Lack of information about the target model is challenging~\cite{chen2020devil}. As strict black-box inference has little knowledge about the target model's performance, it is hard for shadow models to mimic a target model. (ii) User-level inference requires a higher level of robustness than record-level inference. Unlike record-level, user-level inference needs to consider the speaker's voice characteristics. (iii) ASR systems are complicated due to their learning architectures~\cite{chen2020devil}, causing membership inference with shadow models to be computationally resource and time consuming. Finally, time-series audio data is significantly more complex than textual data, resulting in varied feature patterns~\cite{shokoohi2015discovery, du2019sirenattack}.

In this paper, we design and evaluate our audio auditor to help users determine whether their audio records have been used to train an ASR model without their consent. We investigate two types of targeted ASR models: a hybrid ASR system and an end-to-end ASR system. With an audio signal input, both of the models transcribe speech into written text. The auditor audits the target model with an intent via strict black-box access to infer user-level membership. The auditor will behave differently depending on whether audio is transcribed from within its training set or from other datasets. Thus, one can analyze the transcriptions and use the outputs to train a binary classifier as the auditor. As our primary focus is to infer user-level membership, instead of using the rank lists of several top output results, we only use one text output, the user's speed, and the input audio's true transcription while analyzing the transcription outputs (see details in Section~\ref{sec:audit}).

In summary, the main contributions of this paper are as follows:
\begin{enumerate}
    \item We propose the use of user-level membership inference for auditing the ASR model under strict black-box access. With access to the top predicted label only, our audio achieves 78.81\% accuracy. In comparison, the best accuracy for the user-level auditor in text generative models with one top-ranked output is 72.3\%~\cite{song2019auditing}. 
    \item Our auditor is effective in user-level audit. For the user who has audios within the target model's training set, the accuracy of our auditor querying with these recordings can achieve more than 80\%. In addition, only nine queries are needed for each user (regardless of their membership or non-membership) to verify their presence of recordings in the ASR model, at an accuracy of 75.38\%.
    \item Our strict black-box audit methodology is robust to various architectures and pipelines of the ASR model. We investigate the auditor by auditing the ASR model trained with LSTM, RNNs, and GRU algorithms. In addition, two state-of-the-art pipelines in building ASR models are implemented for validation. The overall accuracy of our auditor achieves approximately 70\% across various ASR models on auxiliary and cross-domain datasets. 
    \item We conduct a proof-of-concept test of our auditor on iPhone Siri, under the strict black-box access, achieving an overall accuracy in excess of 80\%. This real-world trial lends evidence to the comprehensive synthetic audit outcomes observed in this paper.
\end{enumerate}

To the best of our knowledge, this is the \textit{first} paper to examine user-level membership inference in the problem space of voice services. We hope the methodology developed in this paper and findings can inform privacy advocates to overhaul IoT privacy.

\section{Background}
\label{sec:bg}
In this section, we overview the automatic speech recognition models and membership inference attacks.

\begin{figure*}[!htbp] 
\begin{subfigure}{.48\textwidth}
  \centering
  \includegraphics[width=1\linewidth]{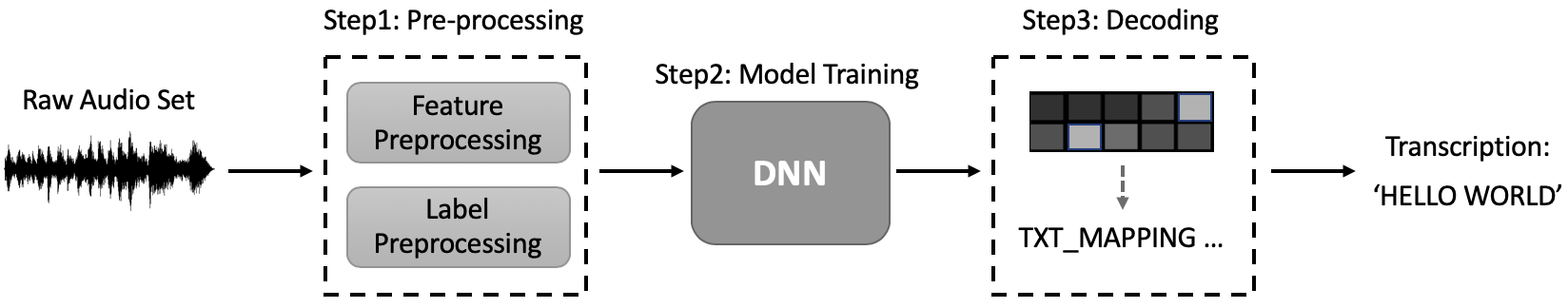}
    \caption{A hybrid ASR system. There are three main steps: (i) the preprocessing step extracts features to represent the raw audio data, (ii) the DNN training step trains the acoustic model and calculates the pseudo-posteriors, and (iii) the decoding step aims to map the predicted symbol combinations to texts and output the transcription results with the highest score.}
    \label{fig:Hybrid_system}
\end{subfigure} 
\qquad
\begin{subfigure}{.48\textwidth}
  \centering
  \includegraphics[width=1\linewidth]{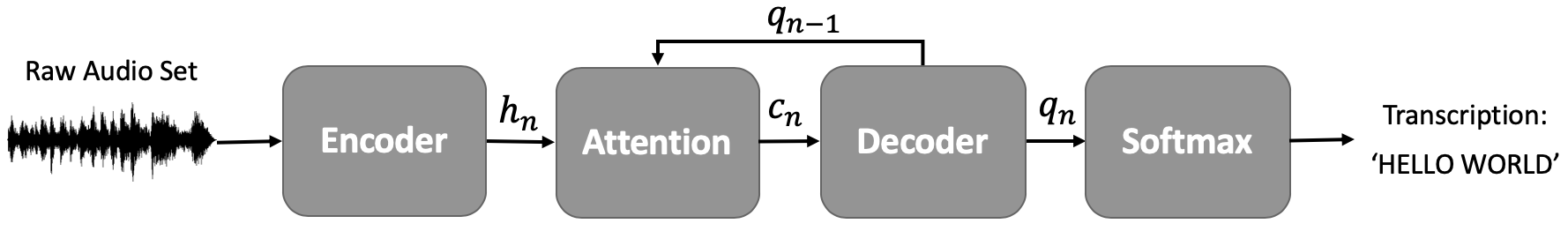}
    \caption{An end-to-end system. There are four main components: (i) the encoder transforms the input audio into a high-level representation $h_{n}$, (ii) the attention mechanism integrates the representation $h_{n}$ with previous decoder output $q_{n-1}$ and gain the context output $c_{n}$, (iii) the decoder decodes the context output $c_{n}$ with previous ground truth as $q_{n}$, and (iv) the Softmax activation like CharDistribution predicts $q_{n}$ auto-regressively and results in final transcription.}
    \label{fig:E2E_system}
\end{subfigure}
\caption{Two state-of-the-art ASR systems}
\label{fig:ASR_model}
\end{figure*}

\subsection{The Automatic Speech Recognition Model}
%%%% Generally describe ASR model
There are two state-of-the-art pipelines used to build the automatic speech recognition (ASR) system, including the typical hybrid ASR systems and end-to-end ASR systems~\cite{perero2018exploring}. To test the robustness of our auditor, we implement both open-source hybrid and end-to-end ASR systems focusing on a speech-to-text task as the target models.

%%%% Hybrid-based ASR model 
\textbf{Hybrid ASR Systems} are mainly DNN-HMM-based acoustic models~\cite{weninger2015speech}. As shown in Fig.~\ref{fig:Hybrid_system}, typically, a hybrid ASR system is composed of a preprocessing step, a model training step, and a decoding step~\cite{schonherr2018adversarial}. During the preprocessing step, features are extracted from the input audio, while the corresponding text is processed as the audio's label. The model training step trains a DNN model to create HMM class posterior probabilities. The decoding step maps these HMM state probabilities to a text sequence. In this work, the hybrid ASR system is built using the pytorch-kaldi speech recognition toolkit~\cite{pytorch-kaldi}. Specifically, feature extraction transforms the audio frame into the frequency domain, as Mel-Frequency Cepstral Coefficients (MFCCs) features. For an additional processing step, feature-space Maximum Likelihood Linear Regression (fMLLR) is used for speaker adaptation. Three popular neural network algorithms are used to build the acoustic model, including Long Short-Term Memory (LSTM), Gated Recurrent Units (GRU), and Recurrent Neural Networks (RNNs). The decoder involves a language model which provides a language probability to re-evaluate the acoustic score. The final transcription output is the sequence of the most suited language with the highest score.

%%%% End-to-end ASR model
\textbf{End-to-End ASR Systems} are attention-based encoder-decoder models~\cite{liu2019adversarial}. Unlike hybrid ASR systems, the end-to-end system predicts sub-word sequences which are converted directly as word sequences. As shown in Fig.~\ref{fig:E2E_system}, the end-to-end system is a unified neural network modeling framework containing four components: an encoder, an attention mechanism, a decoder, and a Softmax layer. The encoder contains feature extraction (i.e., VGG extractor) and a few neural network layers (i.e., BiLSTM layers), which encode the input audio into high-level representations. The location-aware attention mechanism integrates the representation of this time frame with the previous decoder outputs. Then the attention mechanism can output the context vector. The decoder can be a single layer neural network (i.e., an LSTM layer), decoding the current context output with the ground truth of last time frame. Finally, the softmax activation, which can be considered as ``CharDistribution'', predicts several outputs and integrates them into a single sequence as the final transcription.

\subsection{Membership Inference Attack}
%%%% General membership inference
The membership inference attack is considered as a significant privacy threat for machine learning (ML) models~\cite{nasr2018machine}. The attack aims to determine whether a specific data sample is within the target model's training set or not. The attack is driven by the different behaviors of the target model when making predictions on samples within or out of its training set. 

%%%% Previous membership inference against white/black but not strict black
Various membership inference attack methods have been recently proposed. Shokri et al.~\cite{shokri2017membership} train shadow models to constitute the attack model against a target ML model with black-box access. The shadow models mimic the target model's prediction behavior. To improve accuracy, Liu et al.~\cite{liu2019socinf} and Hayes et al.~\cite{hayes2019logan} leverage Generative Adversarial Networks (GAN) to generate shadow models with increasingly similar outputs to the target model. Salem et al.~\cite{salem2019ml} relax the attack assumptions mentioned in the work~\cite{shokri2017membership}, demonstrating that shadow models are not necessary to launch the membership inference attack. Instead, a threshold of the predicted confidence score can be defined to substitute the attack model. Intuitively, a large confidence score indicates the sample as a member of the training set~\cite{song2019privacy}. The attacks mentioned in the work above are all performed on the record level, while Song and Shmatikov~\cite{song2019auditing} study a user-level membership inference attack against text generative models. Instead of using the prediction label along with the confidence score, Song and Shmatikov~\cite{song2019auditing} utilize word's rank list information of several top-ranked predictions as key features to generate the shadow model. Apart from the black-box access, Farokhi and Kaafar~\cite{farokhi2020modelling} model the record-level membership inference attack under the white-box access. 

%%%% Membership inference attack on this paper
Unlike image recognition systems or text generative systems, ASR systems present additional challenges~\cite{chen2020devil}. With strict black-box access, attacks using confidence scores cannot be applied. With limited discriminative power, features can only be extracted from the predicted transcription and its input audio to launch membership inference attacks, i.e., audio auditing in our paper.

\section{Auditing the ASR Models} \label{sec:audit}
\begin{figure*}[ht]
\begin{center}
\centerline{\includegraphics[width=0.95\linewidth]{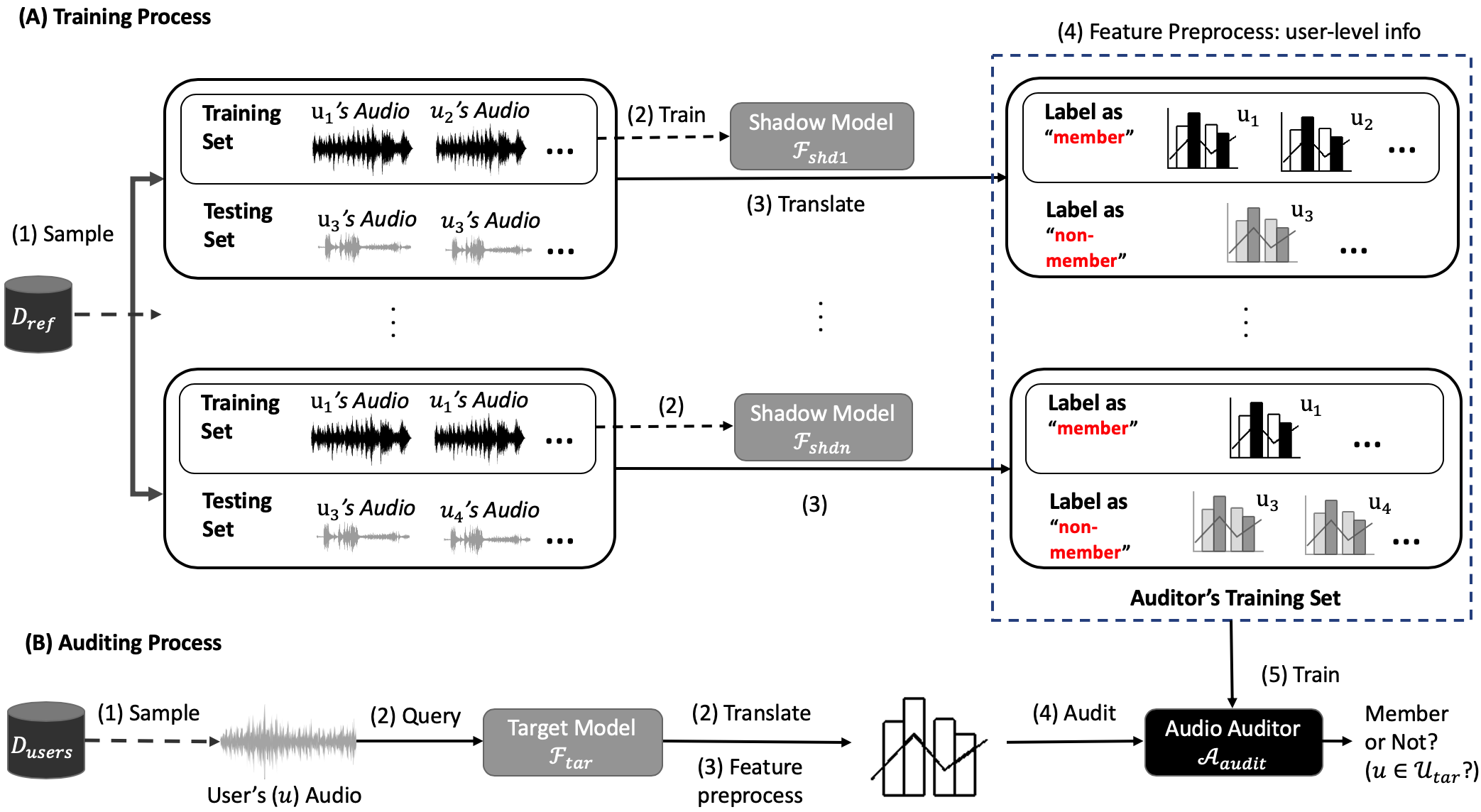}}
\caption{Auditing an ASR model. (i) In the training process, we sample $n$ datasets from the auxiliary reference dataset $D_{ref}$ to build $n$ ($n\geqslant1$) shadow models. Each shadow model dataset $A_{shdi} \sim D_{ref}^{N}, i=1, \ldots, n$ is split to a training set $A_{shdi}^{train}$ and a testing set $A_{shdi}^{test}$. Then we query the shadow model with $A_{shdi}^{test}$ and $A_{shdi}^{train}$ and label their transcriptions as ``\texttt{member}'' or ``\texttt{nonmember}''. Then an audit model can be trained with the outputs of shadow models. (ii) In the auditing process, we randomly sample a particular speaker's ($u$'s) audios $A_{u} \sim D_{users}$ to query our target ASR model. Feature vectors from outputs of the target ASR model can be passed to the audit model to determine whether $u \in U_{tar} \leftarrow A_{tar}$ holds.}
\label{fig:audio_auditor}
\end{center}
\vspace{-0.3in}
\end{figure*}

In this section, we first formalize our objective for auditing ASR models. Secondly, we present how a user-level ASR auditor can be constructed and used to audit the target ASR. Finally, we show how we implement the auditor. 

\subsection{Problem Statement} \label{subsec:problem}
We define \textit{user-level membership inference} as querying a user's data and trying to determine whether any data within the target model's training set belongs to this user.  Even if the queried data is not members of the training set, but data belonging to this user is members in the training set, then this user is regarded as the user-level member of this training set. Let $(x,y) \in \mathbb{X} \times \mathbb{Y}$ denote an audio sample, where $x$ presents the audio component, and $y$ is the actual text of $x$. Assume an ASR model is a function $\mathcal{F}:\mathbb{X}\to\mathbb{Y}$. $\mathcal{F}(x)$ is the model's translated text. The smaller the difference between $\mathcal{F}(x)$ and $y$, the better the ASR model performs. Let $D$ represent a distribution of audio samples. Assume an audio set $A$ is sampled from $D$ of size~$N$ ($A \sim D^{N}$). Let $U$ be the speaker set of $A$ of size~$M$ ($U \leftarrow A$). The ASR model trained with the dataset $A$ is denoted as $\mathcal{F}_{A}$. Let $\mathcal{A}$ represent our auditor, and the user-level auditing process can be formalized as:
\begin{itemize}
    \item A speaker $u$ has $S=\bigcup_{i=1}^{m}(x_{i}, y_{i})$, where $u \leftarrow S$.
    \item Let $Y'=\bigcup_{i=1}^{m}y_{i}'$, when $y_{i}'=\mathcal{F}_{A}(x_{i})$.
    \item Let $``\texttt{member}"=0$ and $``\texttt{nonmember}"=1$.
    \item Set $r=0$ if $u \in U$, or $r=1$ if $u \notin U$.
    \item The auditor successes if $\mathcal{A}(u, S, Y')=r$; otherwise it fails.
\end{itemize}

% \noindent \textbf{Threat Model.} 
Our auditor, as an application of user-level membership inference, checks a speaker's membership of an ASR model's training set. This ASR model is considered as the target model. To closely mirror the real world, we query the target model with strict black-box access. The model only outputs a possible text sequence as its transcription when submitting an audio sample to the target model. This setting reflects the reality, as the auditor may not know this transcription's posterior probabilities or other possible transcriptions. Additionally, any information about the target model is unknown, including the model's parameters, algorithms used to build the model, and the model's architecture. To evaluate our auditor, we develop our target ASR model $\mathcal{F}_{tar}$ using an audio set $A_{tar}$ with two popular pipelines --- hybrid ASR model and end-to-end ASR model --- to represent the ASR model in the real world. As described in Section~\ref{sec:bg}, the hybrid ASR model and the end-to-end ASR model translate the audio in different manners. Under the strict black-box access, the auditor only knows query audio records of a particular user $u$ and its corresponding output transcription. The goal of the auditor is to build a binary classifier $\mathcal{A}_{audit}$ to discriminate whether this user is the member of the user set in which their audio records have been used as target model's training data ($u \in U_{tar}$, $U_{tar} \leftarrow A_{tar}$).

\subsection{Overview of the Proposed Audio Auditor} \label{subsec:overview}
The nature of membership inference~\cite{shokri2017membership} is to learn the difference of a model fed with its actual training samples and other samples. User-level membership inference, like its record-level variant, requires higher robustness. Apart from the disparity of the target model's performance on record-level, our auditor needs to consider the speaker's characteristics as well. Since the posterior probabilities (or confidence scores) are not part of the outputs, shadow models are necessary to audit the ASR model. 

Fig.~\ref{fig:audio_auditor} depicts a workflow of our audio auditor auditing an ASR model. Generally, there are two processes, i.e., training and auditing. The former process is to build a binary classifier as a user-level membership auditor~$\mathcal{A}_{audit}$ using a supervised learning algorithm. The latter uses this auditor to audit an ASR model $\mathcal{F}_{tar}$ by querying a few audios spoken by one user $u$. In Section~\ref{subsec:num_per_user}, we show that only a small number of audios per user can determine whether $u \in U_{tar}$ or $u \notin U_{tar}$. Furthermore, a small number of users used to train the auditor is sufficient to provide a satisfying result.

\noindent \textbf{Training Process.}
The primary task in the training process is to build up shadow models of high quality. Shadow models, mimicking the target model's behaviors, try to infer the targeted ASR model's decision boundary. Due to strict black-box access, a good quality shadow model performs with an approximate testing accuracy as the target model. We randomly sample $n$ datasets from the auxiliary reference dataset $D_{ref}$ as $A_{shd1}, \ldots, A_{shdn}$ to build $n$ shadow models. Each shadow model's audio dataset $A_{shdi}, i=1, \ldots, n$ is split to a training set $A_{shdi}^{train}$ and a testing set $A_{shdi}^{test}$. To build up the ground truth for auditing, we query the shadow model with $A_{shdi}^{train}$ and $A_{shdi}^{test}$. Assume a user's audio set $A_{u}$ is sampled from users' audio sets $D_{users}$. According to the user-level membership inference definition, the outputs from the audio $A_{u} \in A_{shdi}^{test}$ where its speaker $u \notin U_{shdi}^{train}$ are labeled as ``\texttt{nonmember}''. Otherwise, the outputs translated from the audio $A_{u} \in A_{shdi}^{train}$ and from the audio $A_{u} \in A_{shdi}^{test}$ where its speaker $u \in U_{shdi}^{train}$ are all labeled as ``\texttt{member}''. Herein, $U_{shdi}^{train} \leftarrow A_{shdi}^{train}$. To simplify the experiment, for each shadow model, training samples are disjoint from testing samples ($A_{shdi}^{train} \cap A_{shdi}^{test} = \emptyset$). Their user sets are disjoint as well ($U_{shdi}^{train} \cap U_{shdi}^{test} = \emptyset$). With some feature extraction (noted below), those labeled records are gathered as the auditor model's training set.

Feature extraction is another essential task in the training process. Under the strict black-box access, features are extracted from the input audio, ground truth transcription, and the predicted transcription. As a user-level membership inferrer, our auditor needs to learn the information about the target model's performance and the speaker's characteristics. Comparing the ground truth transcription and the output transcription, the \textbf{similarity score} is the first feature to represent the ASR model's performance. To compute the two transcriptions' similarity score, the GloVe model~\cite{pennington2014glove} is used to learn the vector space representation of these two transcriptions. Then the cosine similarity distance is calculated as the two transcriptions' similarity score. Additionally, the input \textbf{audio frame length} and the \textbf{speaking speed} are selected as two features to present the speaker's characteristics. Because a user almost always provides several audios to train the ASR model, statistical calculation is applied to the three features above, including sum, maximum, minimum, average, median, standard deviation, and variance. After the feature extraction, all user-level records are gathered with labels to train an auditor model using a supervised learning algorithm.

To test the quality of the feature set above, we trained an auditor with 500 user-level samples using the Random Forest (RF) algorithm. By randomly selecting 500 samples 100 times, we achieve an average accuracy result over 60\%. Apart from the three aforementioned features, two additional features are added to capture more variations in the model's performance, including \textbf{missing characters} and \textbf{extra characters} obtained from the transcriptions. For example, if \textit{(truth transcription, predicted transcription) = (THAT IS KAFFAR'S KNIFE, THAT IS CALF OUR'S KNIFE)}, then \textit{(missing characters, extra characters) = (KFA, CL OU)}. Herein, the blank character in the extra characters means that one word was mistranslated as two words. With these two extra features, a total of five features are extracted from record-level samples: \textit{similarity score, missing characters, extra characters, frame length,} and \textit{speed}. The record-level samples are transformed into user-level samples using statistical calculation as previously described. We compare the performance of two auditors trained with the two feature sets. We also consider adding 13 Mel-Frequency Cepstral Coefficients (MFCCs) as the additional audio-specific feature set to accentuate each user's records with the average statistics. As seen in Table~\ref{tab:verify_features}, the statistical feature set with 5-tuple is the best choice with approximately 80\% accuracy, while the results with additional audio-specific features are similar, but trail by one percentage. Thus, we proceed with five statistical features to represent each user as the outcome of the feature extraction step.

\begin{table}[t]
\centering
\captionsetup{justification=raggedright, singlelinecheck=false}
\caption{The audit model's performance when selecting either 3 features, 5 features, or 5 features with MFCCs for each audio's query.}
\label{tab:verify_features}
\resizebox{\linewidth}{!}{%
\begin{tabular}{l|c|c|c|c}
 & F1-score & Precision & Recall & Accuracy \\ \hline\hline
Feature\_Set3 & 63.89\% & 68.48\% & 60.84\% & 61.13\% \\ \hline
Feature\_Set5 & 81.66\% & 81.40\% & 82.22\% & 78.81\% \\ \hline
Feature\_Set5 + MFCCs &  81.01\% & 79.72\% & 82.52\% & 77.82\% \\
\end{tabular}%
}
\end{table}

\noindent \textbf{Auditing Process.}
After training an auditor model, we randomly sample a particular speaker's ($u's$) audios~$A_{u}$ from $D_{users}$ to query our target ASR model. With the same feature extraction, the outputs can be passed to the auditor model to determine whether this speaker $u \in U_{tar}$. We assume that our target model's dataset $D_{tar}$ is disjoint from the auxiliary reference dataset $D_{ref}$ ($D_{tar} \cap D_{ref} = \emptyset$). In addition, $U_{ref}$ and $U_{tar}$ are also disjoint ($U_{tar} \cap U_{ref} = \emptyset$). For each user, as we will show, only a limited number of audios are needed to query the target model and complete the whole auditing phase.

\subsection{Implementation}
\label{sec:exp}
In these experiments, we take audios from LibriSpeech~\cite{panayotov2015librispeech}, TIMIT~\cite{garofolo1993timit}, and TED-LIUM~\cite{rousseau2012ted} to build our target ASR model and the shadow model. Detailed information about the speech corpora and model architectures can be found in the Appendix. Since the LibriSpeech corpus has the largest audio sets, we primarily source records from LibriSpeech to build our shadow models.

% \subsubsection{Datasets} \label{subsec:dataset}
% The \noindent\textbf{LibriSpeech} speech corpus (LibriSpeech) contains 1000 hours of speech audios from audiobooks which are part of the LibriVox project~\cite{panayotov2015librispeech}. This corpus is famous in training and evaluating speech recognition systems. At least 1,500 speakers have contributed their voice to this corpus. We use 100 hours of clean speech data with 29,877 recordings to train and test our target model. 360 hours of clean speech data, including 105,293 recordings are used for training and testing the shadow models. Additionally, there are 500 hours of noisy data used to train the ASR model and to test our auditor's performance in noisy environmnet.

% The \noindent\textbf{TIMIT} speech corpus (TIMIT) is another famous speech corpus used to build ASR systems. This corpus recorded audios from 630 speakers in different areas of the United States, totaling 6,300 sentences \cite{garofolo1993timit}. In this work, we use all this data to train and test a target ASR model, and then audit this model with our auditor.

% The \noindent\textbf{TED-LIUM} speech corpus (TED) collected audios based on TED Talks for ASR development \cite{rousseau2012ted}. This corpus was built from the TED talks of the IWSLT 2011 Evaluation Campaign. There are 118 hours of speech with corresponding transcripts.

% \subsubsection{Target Model}
\noindent \textbf{Target Model.} 
Our target model is a speech-to-text ASR model. The inputs are a set of audio files with their corresponding transcriptions as labels, while the outputs are the transcribed sequential texts. To simulate most of the current ASR models in the real world, we created a state-of-the-art hybrid ASR model~\cite{schonherr2018adversarial} using the PyTorch-Kaldi Speech Recognition Toolkit~\cite{pytorch-kaldi} and an end-to-end ASR model using the Pytorch implementation \cite{liu2019adversarial}. In the preprocessing step, fMLLR features were used to train the ASR model with 24 training epochs. Then, we trained an ASR model using a deep neural network with four hidden layers and one Softmax layer. We experimentally tuned the batch size, learning rate and optimization function to gain a model with better ASR performance. To mimic the ASR model in the wild, we tuned the parameters until the training accuracy exceeded 80\%, similar to the results shown in~\cite{perero2018exploring, liu2019adversarial}. Additionally, to better contextualize our audit results, we report the overfitting level of the ASR models, defined as the difference between the predictions' Word Error Rate (WER) on the training set and the testing set ($Overfitting = WER_{train}-WER_{test}$).

\begin{table}[t]
\captionsetup{justification=raggedright, singlelinecheck=false}
\caption{The audit model's performance trained with different algorithms.}
\label{tab:verify_auditor_algorithm}
\centering
\resizebox{0.85\linewidth}{!}{
\begin{tabular}{l|c|c|c|c}
 & F1-score & Precision & Recall & Accuracy \\ \hline\hline
DT & 68.67\% & 70.62\% & 67.29\% & 64.97\% \\ \hline
RF & 81.66\% & 81.40\% & 82.22\% & 78.81\% \\ \hline
$3$-NN & 58.62\% & 64.49\% & 54.69\% & 56.16\% \\ \hline
NB & 34.42\% & 93.55\% & 21.09\% & 53.96\% \\
\end{tabular}%
}
\end{table}

\section{Experimental Evaluation and Results}
\label{sec:eva}

The goal of this work is to develop an auditor for users to inspect whether their audio information is used without consent by ASR models or not. We mainly focus on the evaluation of the auditor, especially in terms of its effectiveness, efficiency, and robustness. As such, we pose the following research questions.
\begin{itemize}
    \item \textit{The effectiveness of the auditor.} We train our auditor using different ML algorithms and select one with the best performance. How does the auditor perform with different sizes of training sets? How does it perform in the real-world scenario, such as auditing iPhone Siri?
    \item \textit{The efficiency of the auditor.} How many pieces of audios does a user need for querying the ASR model and the auditor to gain a satisfying result?
    \item \textit{The data transferability of the auditor.} If the data distribution of the target ASR model's training set is different from that of the auditor, is there any effect on the auditor's performance? If there is a negative effect on the auditor, is there any approach to mitigate it?
    \item \textit{The robustness of the auditor.} How does the auditor perform when auditing the ASR model built with different architectures and pipelines? How does an an auditor perform when a user queries the auditor with audios recorded in a noisy environment (i.e., noisy queries)?
\end{itemize}

\subsection{Effect of the ML Algorithm Choice for the Auditor} 
We evaluate our audio auditor as a user-level membership inference model against the target ASR system. This inference model is posed as a binary classification problem, which can be trained with a supervised ML algorithm. We first consider the effect of different training algorithms on our auditor performance.

To test the effect of different algorithms on our audit methodology, we need to train one shadow ASR model for training the auditor and one target ASR model for the auditor's auditing phase. We assume the target ASR model is a hybrid ASR system whose acoustic model is trained with a four-layer LSTM network. The training set used for the target ASR model is 100 hours of clean audio sampled from the LibriSpeech corpus~\cite{panayotov2015librispeech}. Additionally, the shadow model is trained using a hybrid ASR structure where GRU network is used to build its acoustic model. According to our audit methodology demonstrated in Fig.~\ref{fig:audio_auditor}, we observe the various performance of the audio auditor trained with four popular supervised ML algorithms listing as Decision Tree (DT), Random Forest (RF), $k$-Nearest Neighbor where $k=3$ ($3$-NN), and Naive Bayes (NB). After feature extraction, 500 users' samples from the shadow model's query results are randomly selected as the auditor's training set. To avoid potential bias in the auditor, the number of ``member'' samples and the number of ``nonmember'' samples are equal in all training set splits ($\#\{u \in U_{shd}\}=\#\{u \notin U_{shd}\}$). An additional step taken to eliminate bias, is that each experimental configuration is repeated 100 times. Their average result is reported as the respective auditor's final performance in Table~\ref{tab:verify_auditor_algorithm}.

As shown in Table~\ref{tab:verify_auditor_algorithm}, our four metrics of accuracy, precision, recall, and F1-score are used to evaluate the audio auditor. In general, the RF auditor achieves the best performance compared to the other algorithms. Specifically, the accuracy approaches 80\%, with the other three metrics also exceeding 80\%. We note that all auditors' accuracy results exceed the random guess (50\%). Aside from the RF and DT auditors, the auditor with other ML algorithms behaves significantly differently in terms of precision and recall, where the gaps of the two metrics are above 10\%. The reason is in part due to the difficulty in distinguishing the ``member'' and ``nonmember'' as a user's audios are all transcribed well at a low speed with short sentences. Tree-based algorithms, with right sequences of conditions, may be more suitable to discriminate the membership. We regard the RF construction of the auditor as being the most successful; as such, RF is the chosen audio auditor algorithm for the remaining experiments.

\subsection{Effect of the Number of Users Used in Training Set of the Auditor} \label{sec:number of users}

\begin{figure}[t] 
\centering
\includegraphics[width=0.75\columnwidth]{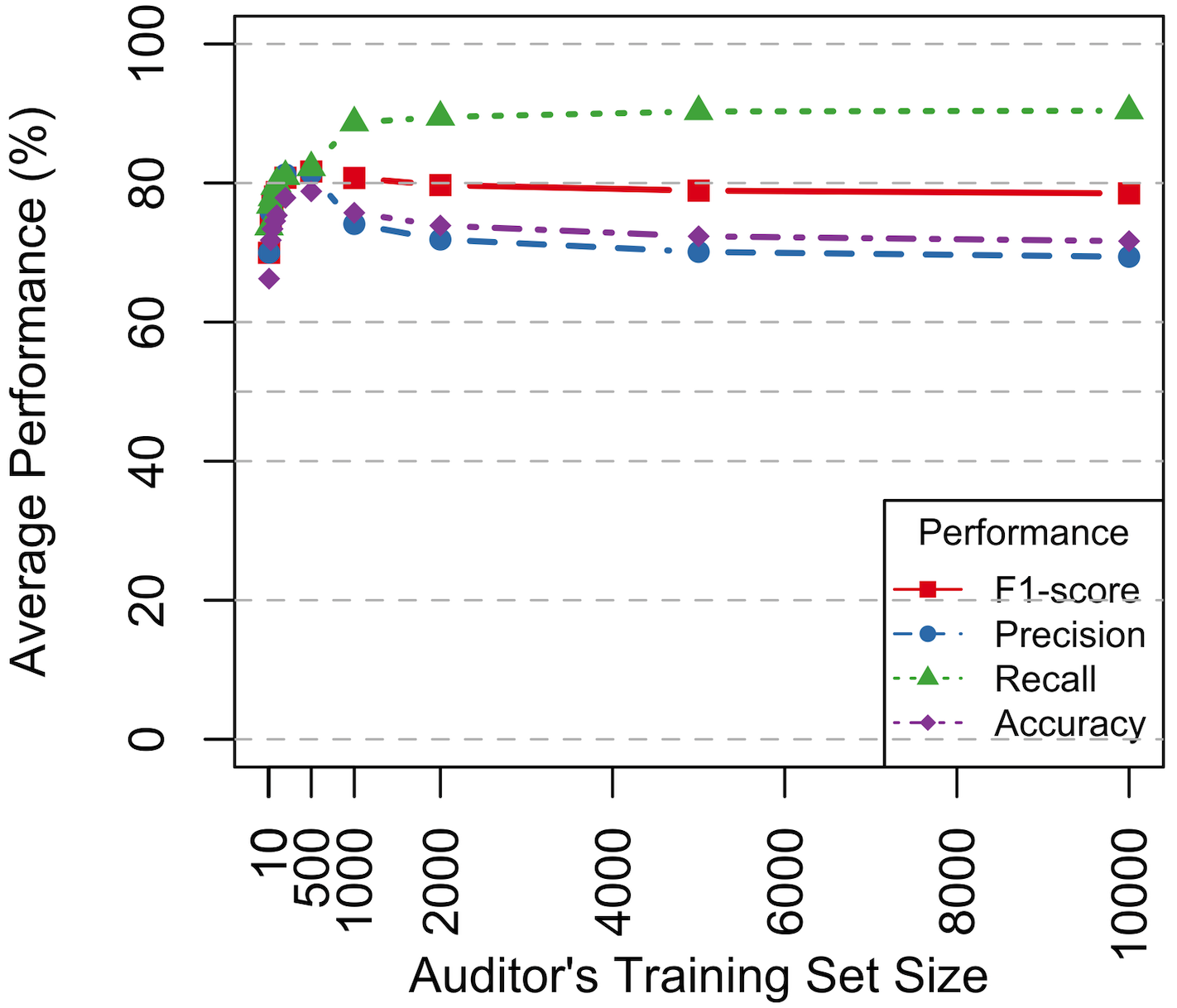}
\caption{Auditor model performance with varied training set size.}
\label{fig:verify_training_size}
\end{figure}

To study the effect of the number of users, we assume that our target model and shadow model are trained using the same architecture (hybrid ASR system). However, due to the strict black-box access to the target model, the shadow model and acoustic model shall be trained using different networks. Specifically, LSTM networks are used to train the acoustic model of the target ASR system, while a GRU network is used for the shadow model. As depicted in Fig.~\ref{fig:verify_training_size}, each training sample $(x_{i}, y_{i})$ is formed by calculating shadow model's querying results for each user $u_{j} \leftarrow \bigcup_{i=1}^{m}(x_{i}, y_{i})$. Herein, we train the audio auditor with a varying number of users $(j=1,...,M)$. The amount of users $M$ we considered in the auditor training set is 10, 30, 50, 80, 100, 200, 500, 1,000, 2,000, 5,000, and 10,000.

On the smaller numbers of users in the auditor's training set, from Fig.~\ref{fig:verify_training_size}, we observe a rapid increase in performance with an increasing number of users. Herein, the average accuracy of the auditor is 66.24\% initially, reaching 78.81\% when the training set size is 500 users. From 500 users, the accuracy decreases then plateaus. Overall, the accuracy is better than the random guess baseline of 50\% for all algorithms. Aside from accuracy, the precision increases from 69.99\% to 80.40\%; the recall is 73.69\% initially and eventually approaches approximately 90\%; and the F1-score is about 80\% when the training set size exceeds 200. In summary, \textit{we identify the auditor's peak performance when using a relatively small number of users for training.}

\begin{figure}[t] % auditor's UMem_in vs UMem_out
\centering
\includegraphics[width=0.75\columnwidth]{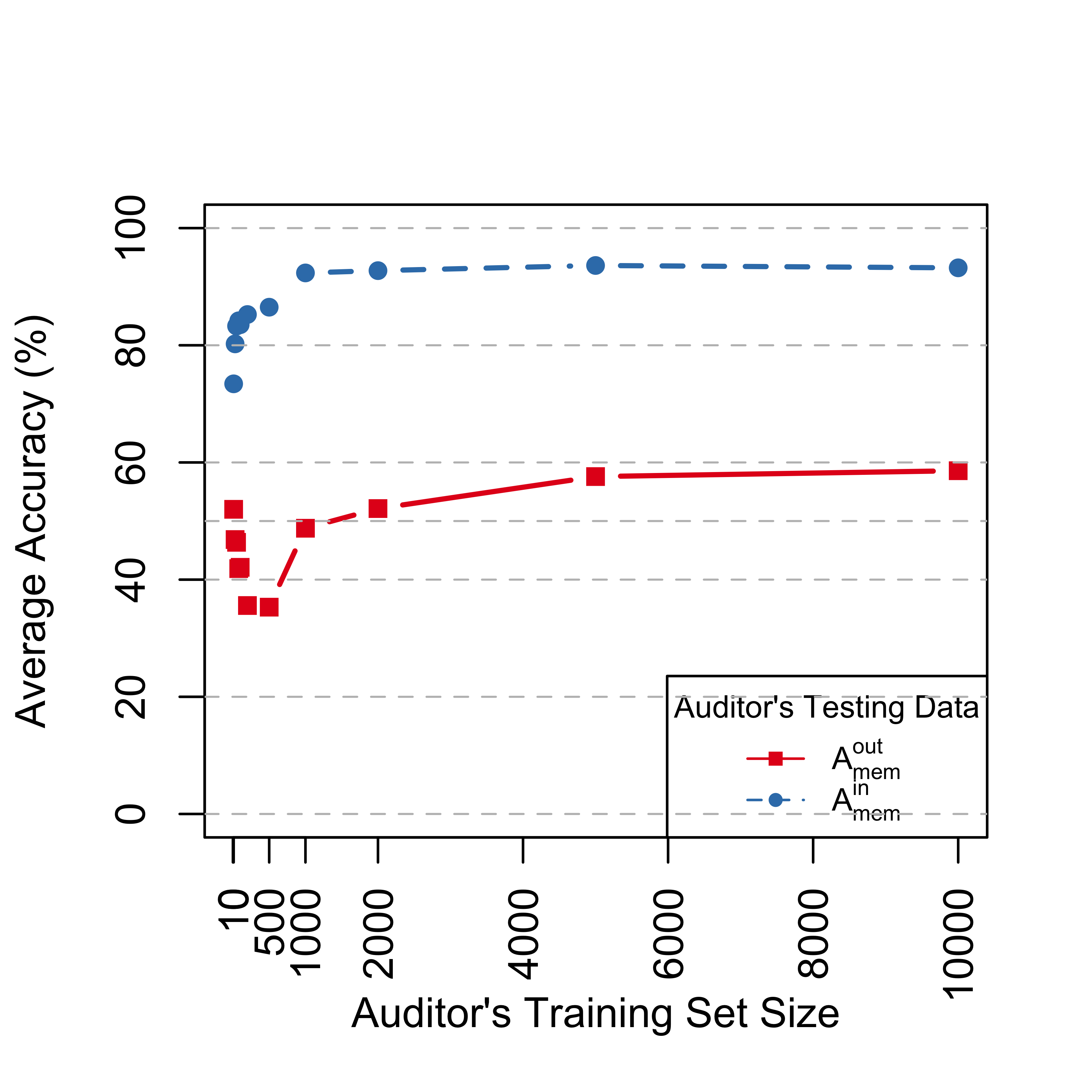}
\caption{Auditor model accuracy on a member user querying with the target model's unseen audios (${A}_{mem}^{out}$) against the performances on the member users only querying with the seen recordings (${A}_{mem}^{in}$).}
\label{fig:various_query_memmix}
\end{figure}

Recall the definition of user-level membership in Section~\ref{subsec:problem}. We further consider two extreme scenarios in the auditor's testing set. One extreme case is that the auditor's testing set only contains member users querying with the unseen audios (excluded from the target model's training set), henceforth denoted as ${A}_{mem}^{out}$. The other extreme case is an auditor's testing set that only contains member users querying with the seen audios (exclusively from the target model's training set), herein marked as ${A}_{mem}^{in}$. Fig.~\ref{fig:various_query_memmix} reports the accuracy of our auditor on ${A}_{mem}^{in}$ versus ${A}_{mem}^{out}$. If an ASR model were to use a user's recordings as its training samples (${A}_{mem}^{in}$), the auditor can determine the user-level membership with a much higher accuracy, when compared to user queries on the ASR model with ${A}_{mem}^{out}$. Specifically, ${A}_{mem}^{in}$ has a peak accuracy of 93.62\% when the auditor's training set size $M$ is 5,000. Considering the peak performance previously shown in Fig.~\ref{fig:verify_training_size}, auditing with ${A}_{mem}^{in}$ still achieves good accuracy (around 85\%) despite a relatively small training set size. Comparing the results shown in Fig.~\ref{fig:verify_training_size} and Fig.~\ref{fig:various_query_memmix}, we can infer that the larger the auditor's training set size is, the more likely nonmember users are to be misclassified. The auditor's overall performance peak when using a small number of the training set is largely due to the high accuracy of the shadow model. A large number of training samples perhaps contain the large proportion of nonmember users' records whose translation accuracy is similar to the member users'. 

\begin{mdframed}[backgroundcolor=black!10,rightline=false,leftline=false,topline=false,bottomline=false,roundcorner=2mm] 
Overall, it is better for users to choose audios that have a higher likelihood of being contained within the ASR model for audit (for example, the audios once heard by the model).
\end{mdframed}

\subsection{Effect of the Target Model Trained with Different Data Distributions}

\begin{figure*}[th]
  \centering
  \begin{subfigure}[b]{0.32\linewidth}
    \centering\includegraphics[width=\linewidth]{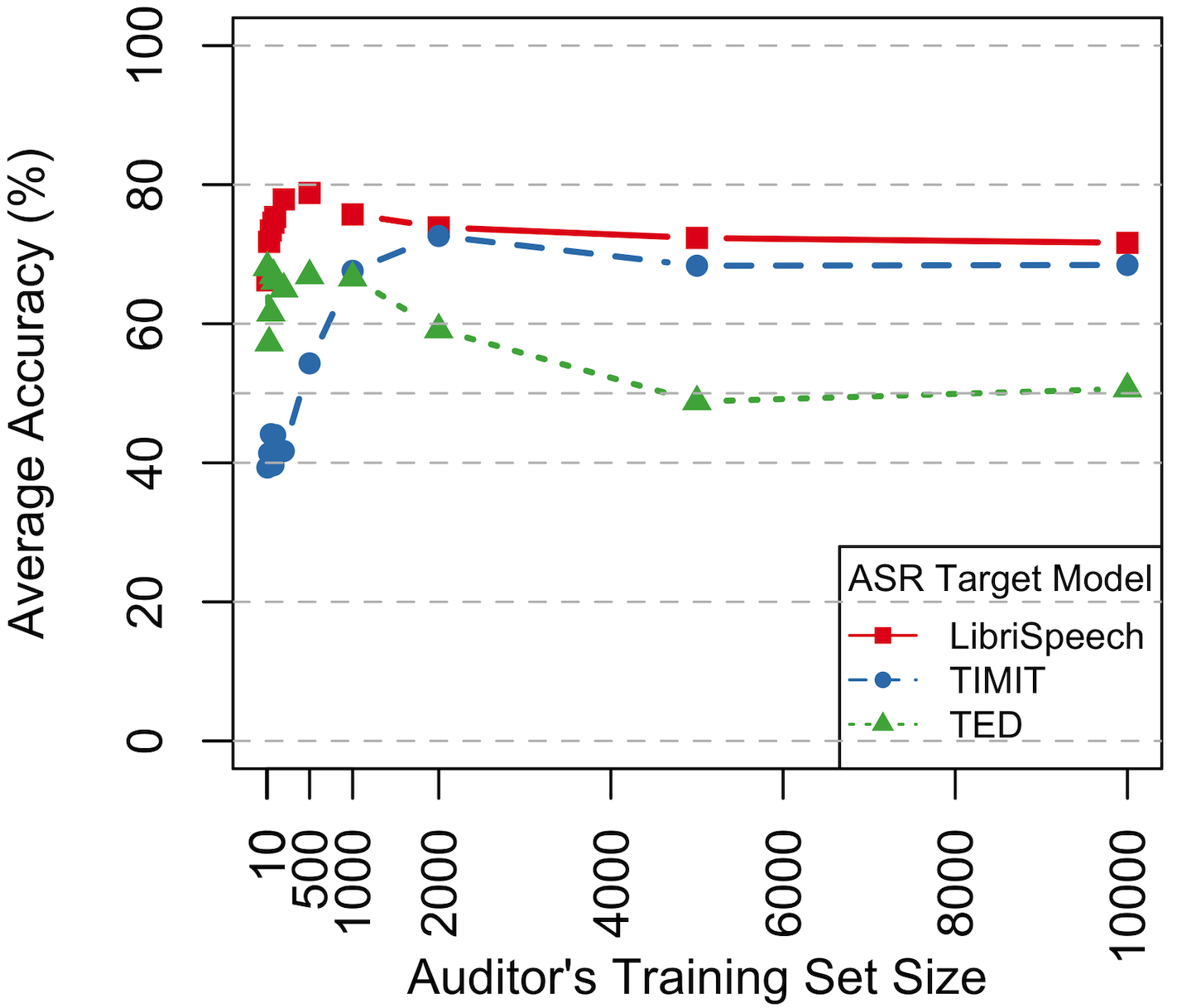}
    \caption{\centering Accuracy}
  \end{subfigure}\hfill%
  \begin{subfigure}[b]{0.32\linewidth}
    \centering\includegraphics[width=\linewidth]{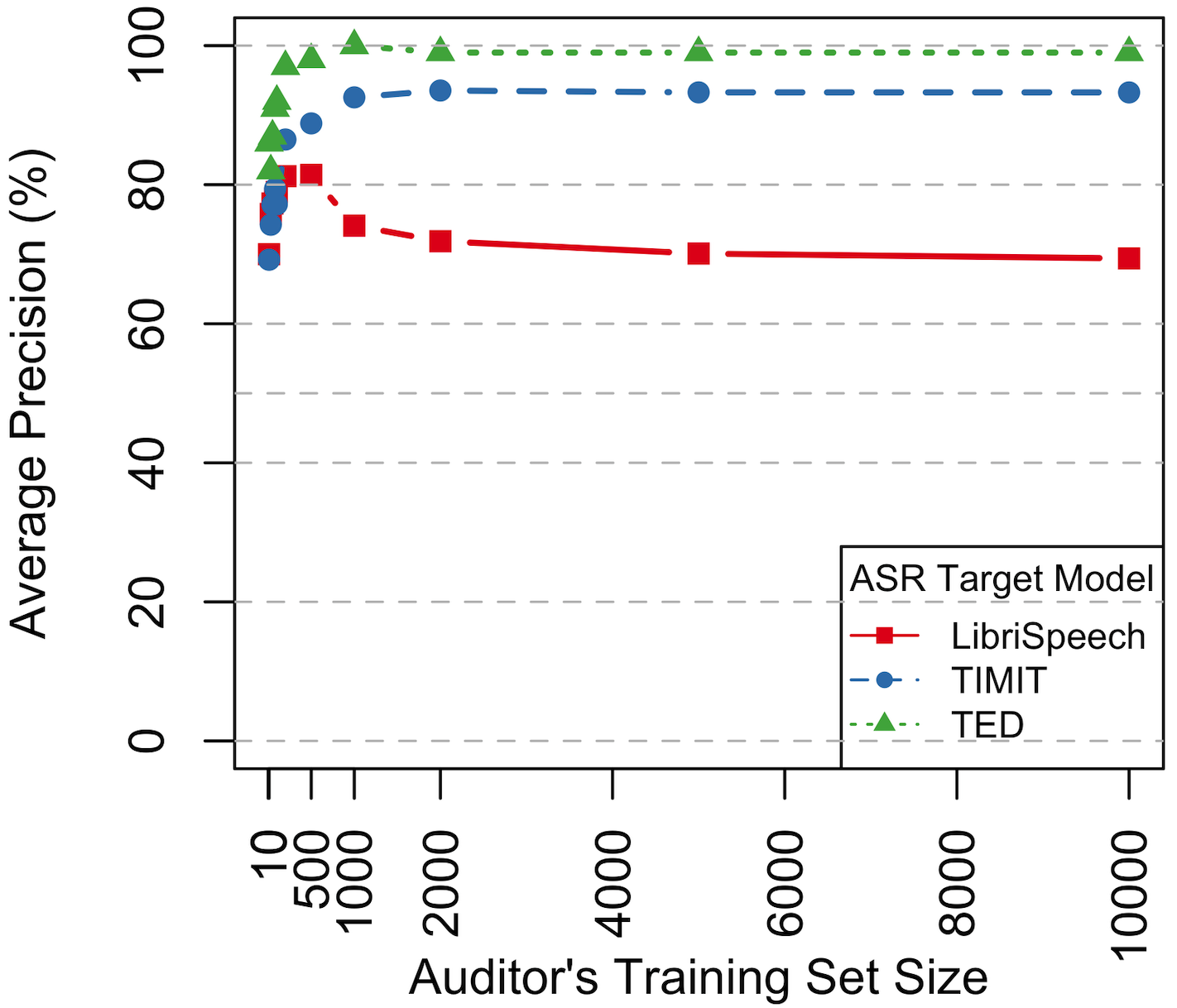}
    \caption{\centering Precision}
  \end{subfigure}\hfill%
  \begin{subfigure}[b]{0.32\linewidth}
    \centering\includegraphics[width=\linewidth]{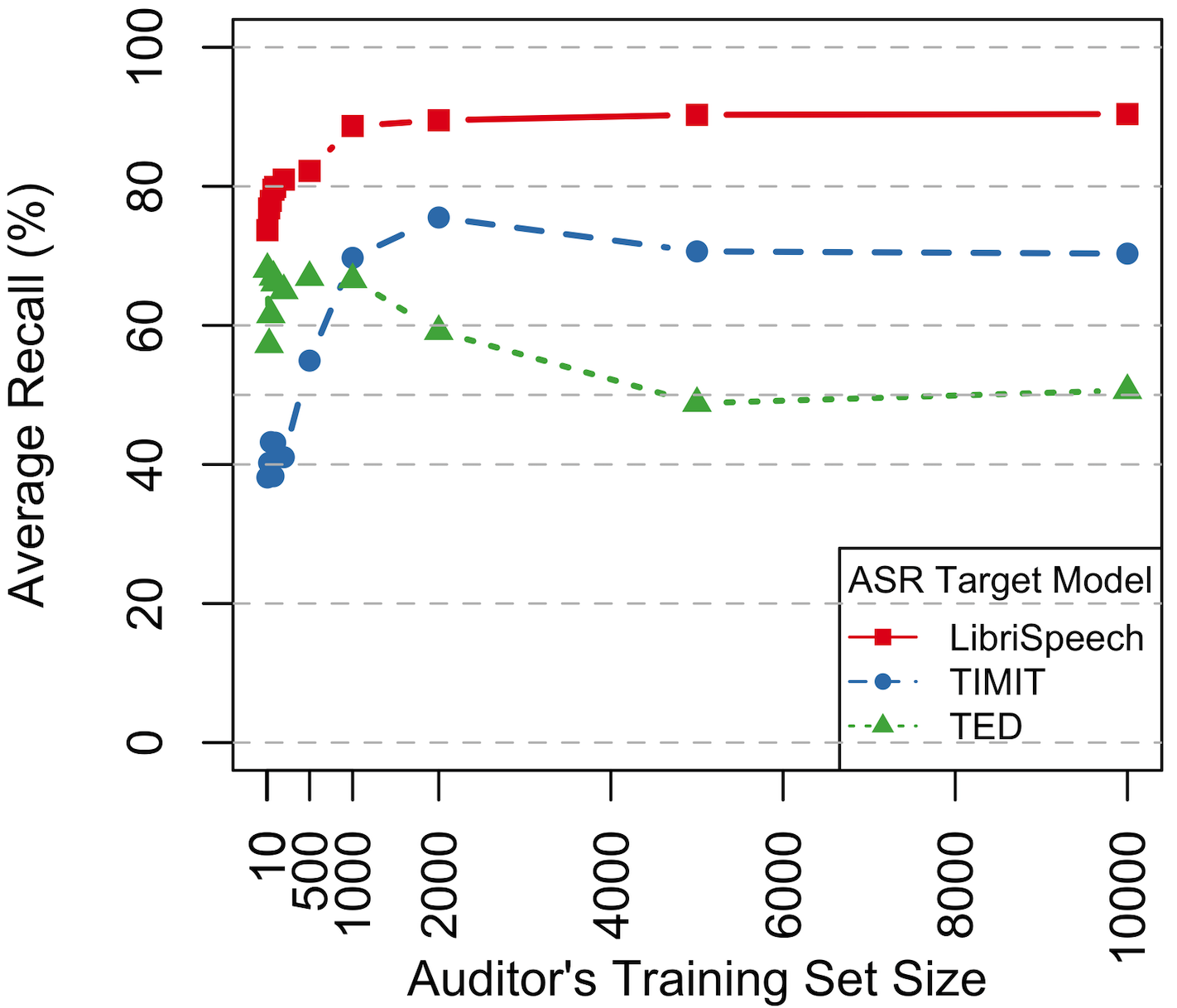}
    \caption{\centering Recall}
  \end{subfigure}
  \caption{The auditor model audits target ASR models trained with training sets of different data distributions. We observe that in regards to accuracy and recall the target model with the same distribution as the auditor performs the best, while the contrary is observed for precision. Nevertheless, the data transferability is well observed with reasonably high metrics for all data distributions.}
  \label{fig:verify_target_data}
\end{figure*}

The previous experiment draws conclusions based on the assumption that the distributions of training sets for the shadow model and the target model are the same. That is, these two sets were sampled from LibriSpeech corpus $D_\mathscr{L}$  ($A_{tar} \sim D_\mathscr{L}$, $A_{shd} \sim D_\mathscr{L}$, $A_{tar} \cap A_{shd} = \emptyset$). Aside from the effects that a changing number of users used to train the auditor, we relax this distribution assumption to evaluate the data transferability of the auditor. To this end, we train one auditor using a training set sampled from LibriSpeech $D_\mathscr{L}$ ($A_{shd} \sim D_\mathscr{L}$). Three different target ASR models are built using data selected from LibriSpeech, TIMIT, and TED, respectively.

Fig.~\ref{fig:verify_target_data} plots the auditor's data transferability on average accuracy, precision, and recall. Once above a certain threshold of the training set size ($\approx10$), the performance of our auditor significantly improves with an increasing number of users data selected as its user-level training samples. Comparing the peak results, the audit of the target model trained with the same data distribution (LibriSpeech) slightly outperforms the audit of target models with different distributions (TIMIT and TED). For instance, the average accuracy of the auditor auditing LibriSpeech data reaches 78.81\% when training set size is 500, while the average audit accuracy of the TIMIT target model peaks at 72.62\% for 2,000 users. Lastly the average audit accuracy of TED target model reaches its maximum of 66.92\% with 500 users. As shown in Fig.~\ref{fig:verify_target_data}, the peaks of precision of the LibriSpeech, TIMIT, and TED target model are 81.40\%, 93.54\%, and 100\%, respectively, opposite of what was observed with accuracy and recall. The observation of the TED target model with extremely high precision and low recall is perhaps due to the dataset's characteristics, where all of the audio clips of TED are long speeches recorded in a noisy environment. 

\begin{mdframed}[backgroundcolor=black!10,rightline=false,leftline=false,topline=false,bottomline=false,roundcorner=2mm] 
In conclusion, our auditor demonstrates satisfying data transferability in general.
\end{mdframed}

\subsection{Effect of the Number of Audio Records per User} \label{subsec:num_per_user}% II: efficiency 
\begin{figure}[t] % queryN_A5 -> queryN_all
  \centering
  \includegraphics[width=0.85\linewidth]{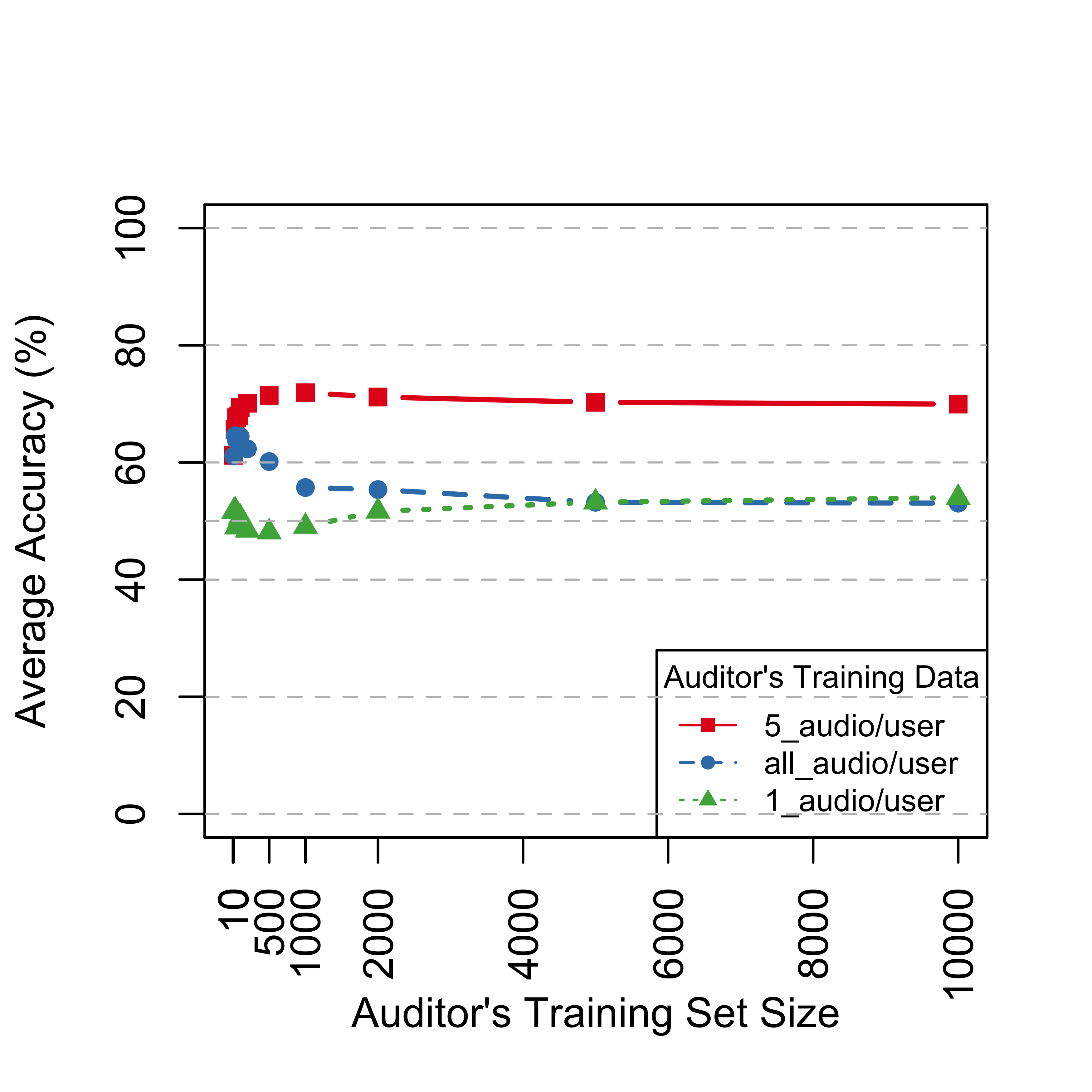}
  \caption{A comparison of average accuracy for one audio, five audios, and all audios per user when training the auditor model with a limited number of audios per user gained in the auditing phase.}
  \label{fig:verify_queryN_A5all}
\end{figure}

\begin{figure}[t]
  \centering
  \includegraphics[width=0.78\linewidth]{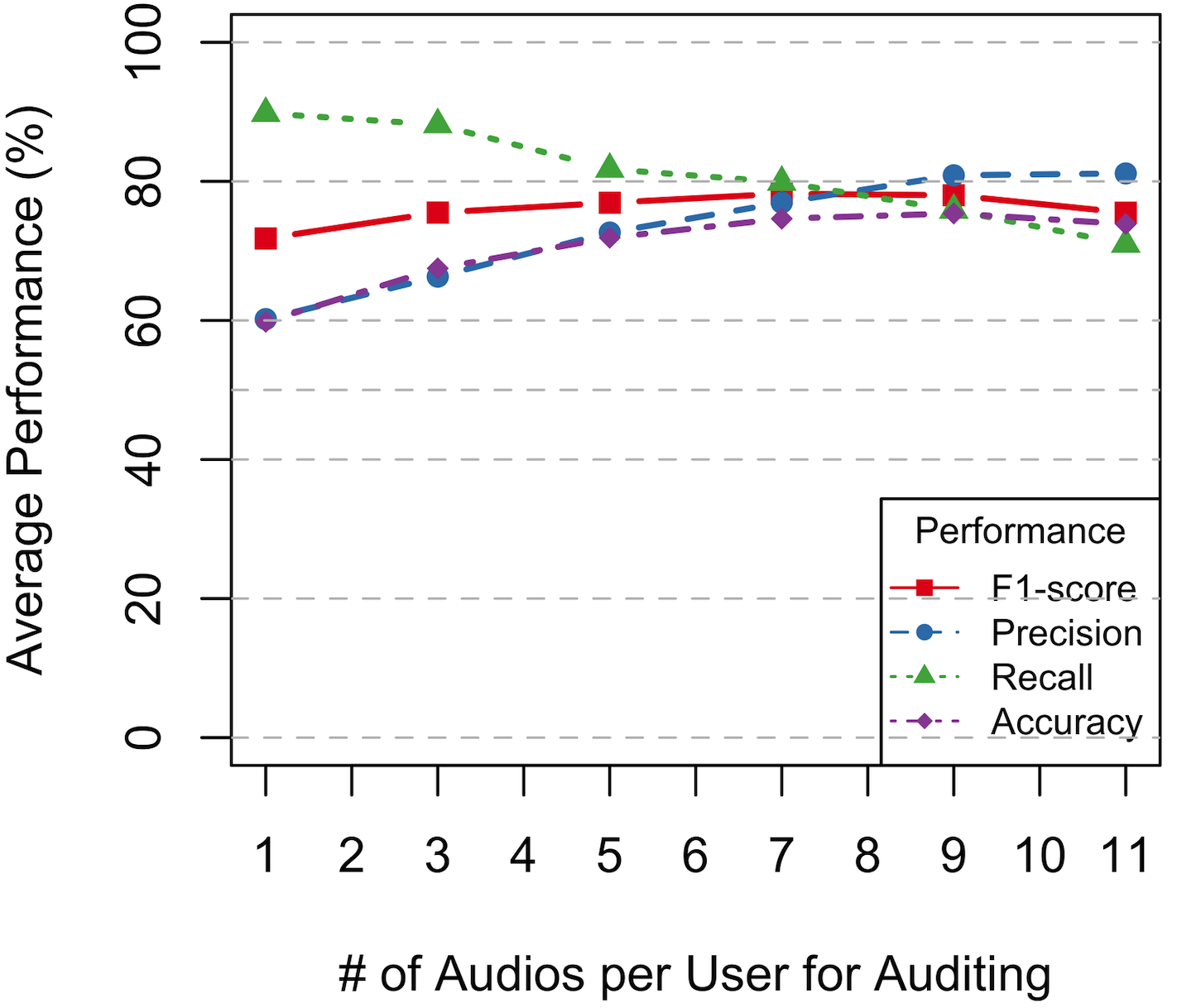}
    \vspace{-0.1in}
  \caption{A varying number of audios used for each speaker when querying an auditor model trained with 5 audios per user.}
  \label{fig:verify_queryN_all}
\end{figure}

The fewer audio samples a speaker is required to submit for their user-level query during the auditing phase, the more convenient it is for users to use the auditor. Additionally, if the auditor can be trained with user-level training samples accumulated from a reduced number of audios per user, both added convenience and the efficiency of feature preprocessing during the auditor's training can be realized. 

\textbf{A limited number of audio samples per user versus a large number of audio samples per user.}
Assuming that each user audits their target ASR model by querying with a limited number of audios, we shall consider whether a small number or a large number of audio samples per user should be collected to train our auditor. Herein, varying the number of audios per user only affects the user-level information learned by the auditor during the training phase. To evaluate this, we have sampled one, five, and all audios per user as the training sets when the querying set uses five audios per user. Fig.~\ref{fig:verify_queryN_A5all} compares the average accuracy of the auditors when their training sets are processed from limits of one audio, five audios, and finally all audios of each user. To set up the five audio auditor's training sets, we randomly select five audios recorded from each user $u_{j} \leftarrow \bigcup_{i=1}^{m=5}(x_{i}, y_{i})$, then translate these audios using the shadow model to produce five transcriptions. Following the feature preprocessing demonstrated in Section~\ref{sec:audit}, user-level information for each user is extracted from these five output transcriptions with their corresponding input audios. The same process is applied to construct the auditor in which the training data consists of one audio per user. To set up the auditor's training set with all the users' samples, we collect all audios spoken by each user and repeat the process mentioned above (average $\bar{m}>62$). Moreover, since the two auditors' settings above rely on randomly selected users, each configuration is repeated 100 times, with users sampled anew, to report the average result free of sampling biases. 

Fig.~\ref{fig:verify_queryN_A5all} demonstrates that the auditor performs best when leveraging five audios per user during the feature preprocessing stage. When a small number of users are present in the training set, the performance of the two auditors is fairly similar, except the auditor trained with one audio per user. For example, when only ten users are randomly selected to train the auditor, the average accuracy of these two auditors are 61.21\% and 61.11\%. When increasing to 30 users in the training set, the average accuracy of the 5-sample and all-sample auditors is 65.65\% and 64.56\%, respectively. However, with more than 30 users in the training set, the auditor trained on five audios per user outperforms that using all audios per user. Specifically, when using five audios per user, the auditor's average accuracy rises to $\approx$70\% with a larger training set size; compared to the auditor using all audios per user, with a degraded accuracy of $\approx$55\%. This is in part owing to the difficulty of accurately characterizing users' audios. In conclusion, despite restrictions on the number of user audio samples when training the auditor, the auditor can achieve superior performance. Consequently, we recommend that the number of audios per user collected for the auditor's training process should be the same for the auditor's querying process.

\textbf{A limited number of audio samples per user while querying the auditor.}
While we have investigated the effect of using a limited number of audios per user to build the training set, we now ask how the auditor performs with a reduced number of audios provided by the user during the querying stage of the audit process, and how many audios each user needs to submit to preserve the performance of this auditor. We assume our auditor has been trained using the training set computed with five audios per user. Fig.~\ref{fig:verify_queryN_all} displays performance trends (accuracy, precision, recall, and F1-score) when a varying number of query audios per user is provided to the target model. We randomly select a user's audios to query the target model by testing  $m=$ 1, 3, 5, 7, 9, or 11 audios per user. As Fig.~\ref{fig:verify_queryN_A5all} reveals that the accuracy results are stable when training set size is large, we conduct our experiments using 10,000 records in the auditor training set. Again, each experiment is repeated 100 times, and the results are all averaged.

Fig.~\ref{fig:verify_queryN_all} illustrates that the auditor performs well with results all above 60\%. Apart from recall, the other three performances trend upwards with an increasing number of audios per user. The scores of the accuracy, precision, and F1-score are approximately 75\%, 81\%, and 78\%, respectively, when each user queries the target model with nine audios, indicating an improvement over the accuracy ($\approx$72\%) we previously observed in Fig.~\ref{fig:verify_training_size}. It appears, for accuracy, when the number of query audios per user grows, the upward trend slows down and even slightly declines. The recall is maximized (89.78\%) with only one audio queried by each user, decreasing to 70.97\% with eleven audios queried for each user. It might happen because the increased number of audio per user does not mean the increased number of users (i.e., testing samples). Since the auditor was trained with five audios per user, the auditor may fail to recognize the user's membership when querying with many more audios.

\begin{mdframed}[backgroundcolor=black!10,rightline=false,leftline=false,topline=false,bottomline=false,roundcorner=2mm] 
Overall, with only a limited number of audios used for audit, e.g.~nine audios per user, our auditor still effectively discriminates a user's membership in the target model's training set.
\end{mdframed}

\subsection{Effect of Training Shadow Models across Different Architectures}% IV: algorithm independent 

\begin{figure*}[th]
  \centering
  \begin{subfigure}[b]{0.32\linewidth}
    \centering\includegraphics[width=\linewidth]{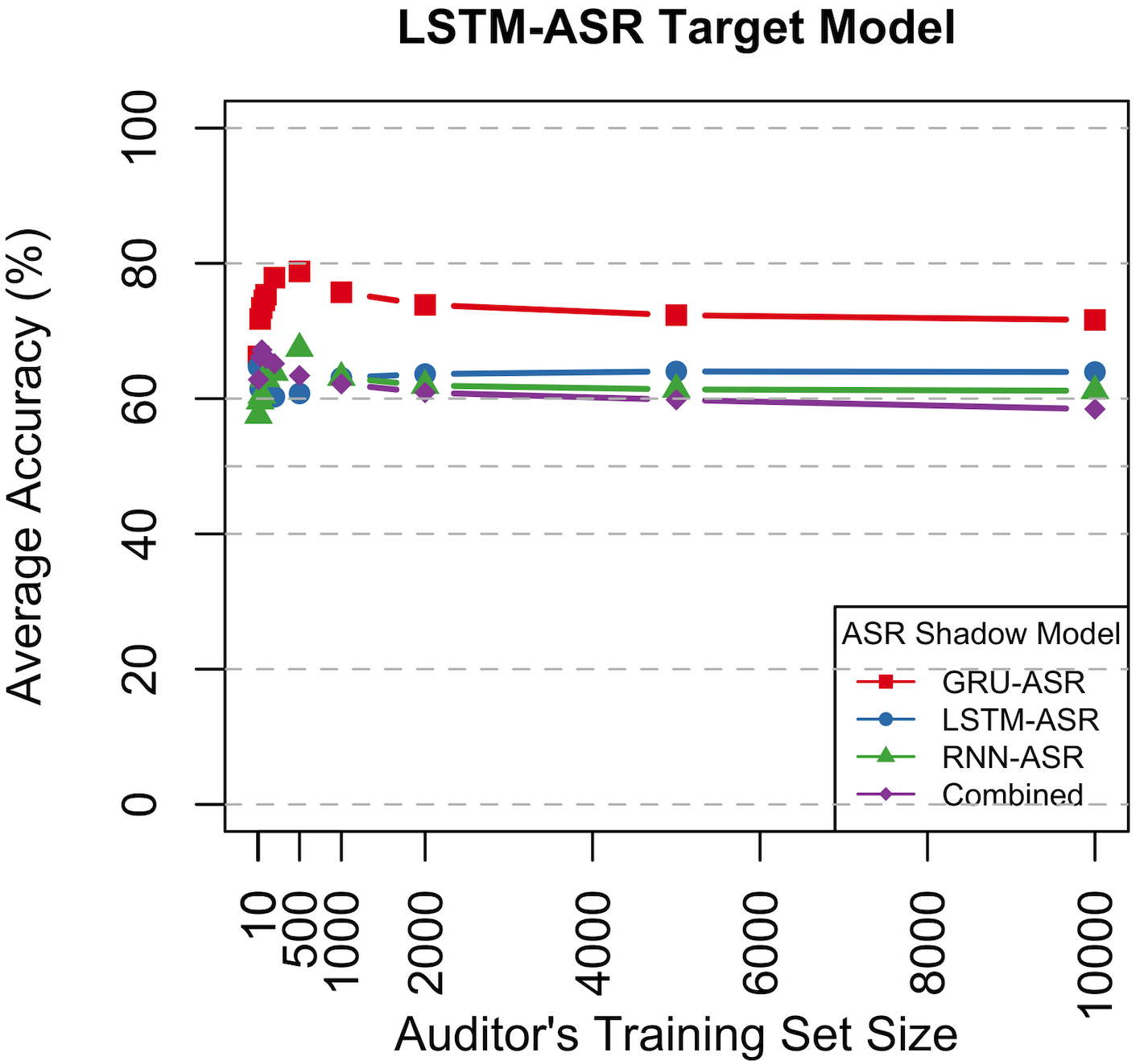}
    \caption{\centering Accuracy}
  \end{subfigure}\hfill%
  \begin{subfigure}[b]{0.32\linewidth}
    \centering\includegraphics[width=\linewidth]{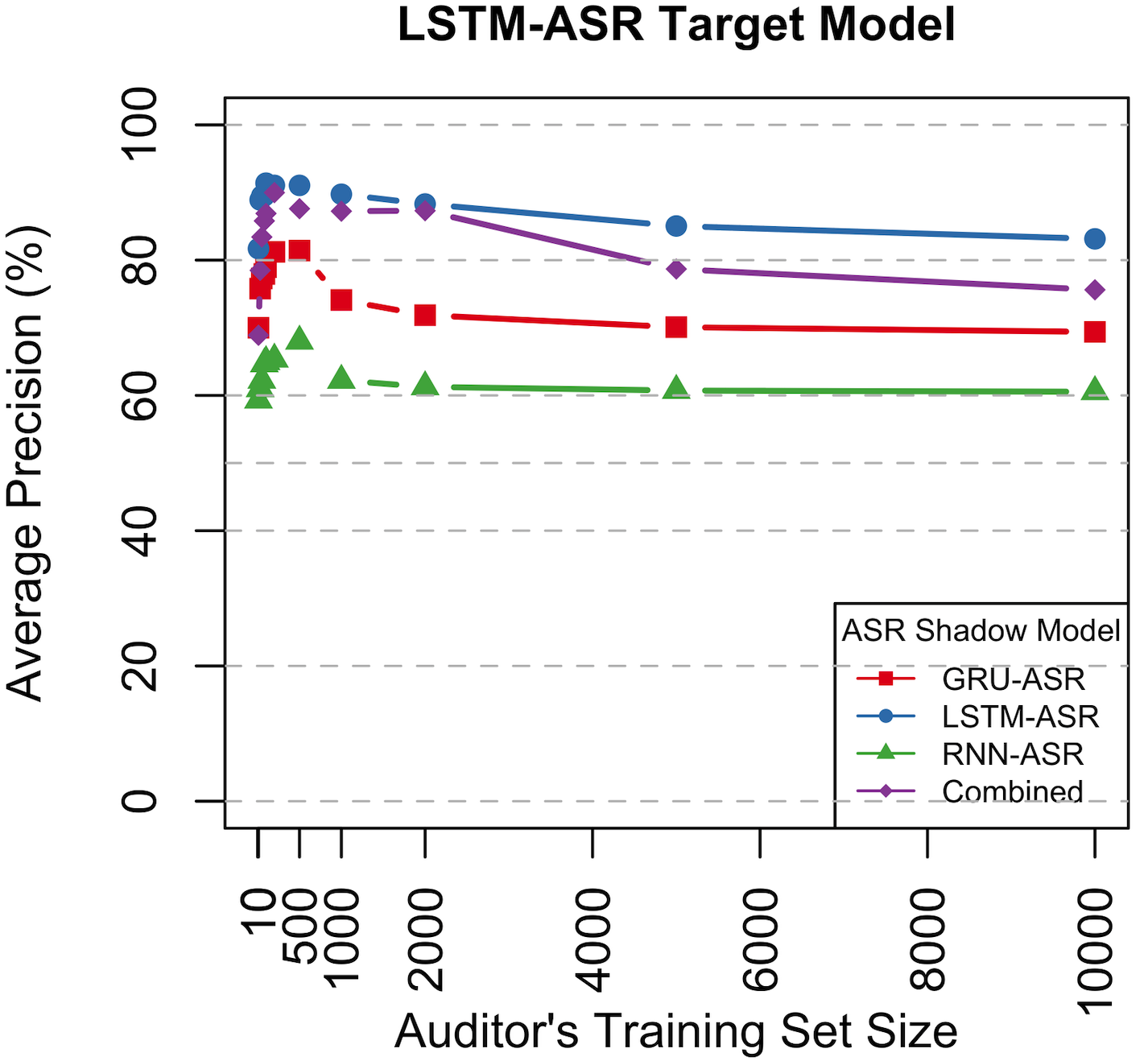}
    \caption{\centering Precision}
  \end{subfigure}\hfill%
  \begin{subfigure}[b]{0.32\linewidth}
    \centering\includegraphics[width=\linewidth]{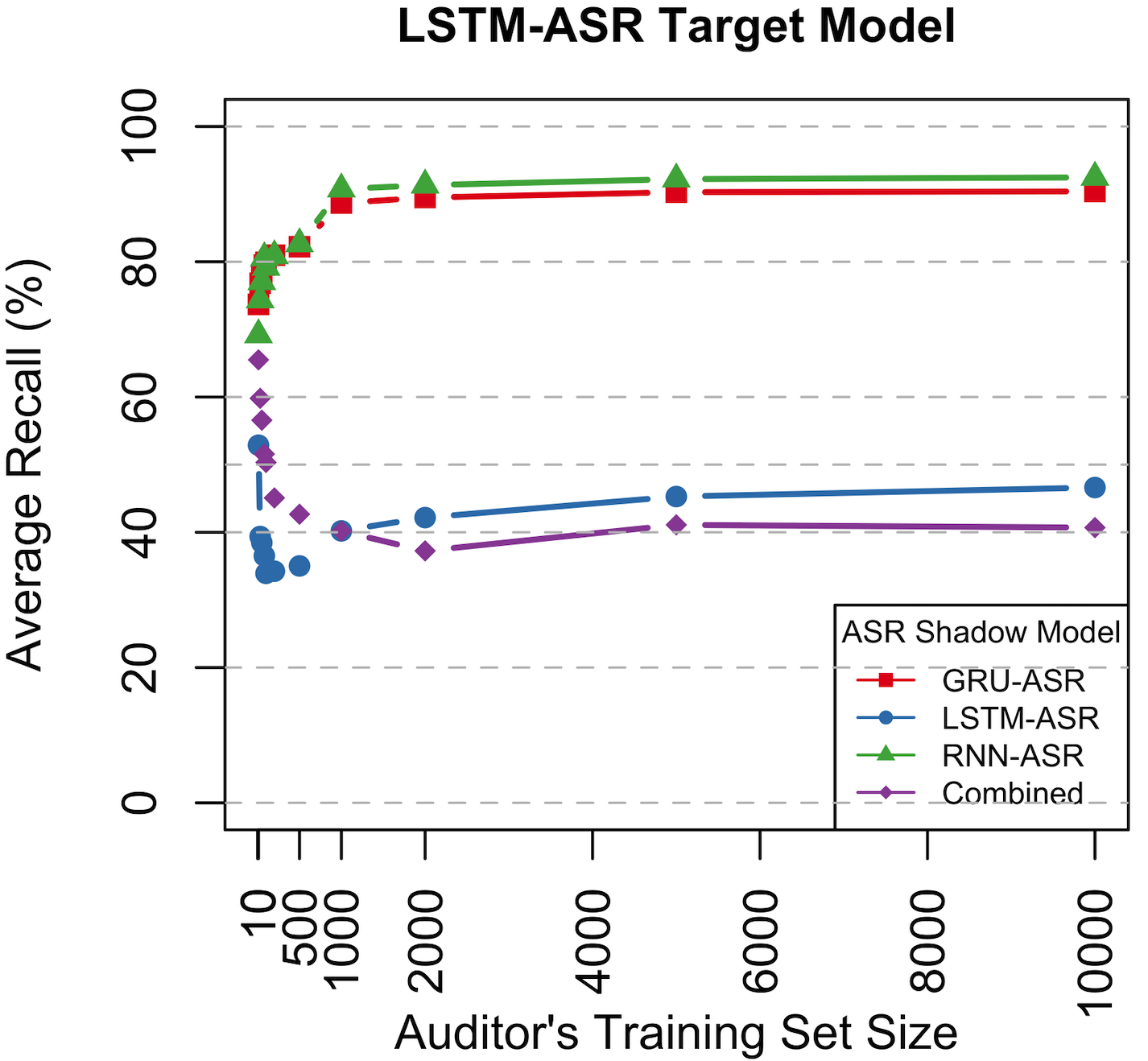}
    \caption{\centering Recall}
  \end{subfigure}
  \caption{Different auditor model performance when trained with different ASR shadow model architectures.}
  \label{fig:verify_train_alg}
\end{figure*}

A shadow model trained with different architectures influences how well it mimics the target model and the performance of the user-level audio auditor. In this subsection, we experiment with different shadow model architectures by training the auditor with information from various network algorithms, like LSTM, RNNs, GRU. If the choice of a shadow model algorithm has a substantial impact on the auditor's performance, we shall seek a method to lessen such an impact. We also seek to evaluate the influence of the combining attack as proposed by Salem et al.~\cite{salem2019ml}, by combining the transcription results from a set of ASR shadow models, instead of one, to construct the auditor's training set. The feature extraction method is demonstrated in Section~\ref{sec:audit}. We refer to this combination as \textit{user-level combining audit}.

\begin{table}[t] % various_ASR_model 
\centering
\caption{Information about ASR models trained with different architectures. ($WER_{train}$: the prediction's WER on the training set; $WER_{test}$: the prediction's WER on the testing set; t: target model; s: shadow model.).}
\label{tab:various_ASR_model}
\resizebox{\linewidth}{!}{
\begin{tabular}{c|c|c|c|c}
ASR Models & Model's Architecture & \multicolumn{1}{p{1.0cm}|}{Dataset Size} & \multicolumn{1}{p{1.4cm}|}{$WER_{train}$} & \multicolumn{1}{p{1.3cm}}{$WER_{test}$} \\ \hline \hline
LSTM-ASR (s) & 4-LSTM layer + Softmax & 360 hrs & 6.48\% & 9.17\% \\ \hline
RNN-ASR (s) & 4-RNN layer +  Softmax & 360 hrs & 9.45\% & 11.09\% \\ \hline
GRU-ASR (s) & 5-GRU layer +  Softmax & 360 hrs & 5.99\% & 8.48\% \\ \hline
LSTM-ASR (t) & 4-LSTM layer +  Softmax & 100 hrs & 5.06\% & 9.08\% \\ 
\end{tabular}
}
\vspace{0.6cm}
\end{table}

To explore the specific impact of architecture, we assume that the acoustic model of the target ASR system is mainly built with the LSTM network (we call this model the LSTM-ASR Target model). We consider three popular algorithms, LSTM, RNNs, and GRU networks, to be prepared for the shadow model's acoustic model. The details of the target and shadow ASR models above are displayed in Table~\ref{tab:various_ASR_model}. Each shadow model is used to translate various audios, with their results processed into the user-level information to train an auditor. Consider this, the shadow model that mainly uses the GRU network structure to train its acoustic model is marked as the GRU-ASR shadow model; its corresponding auditor, named GRU-based auditor, is built using the training set constructed from GRU-ASR shadow model's query results. Our other two auditors have a similar naming convention, an LSTM-based auditor and an RNN-based auditor. Moreover, as demonstrated in Fig.~\ref{fig:audio_auditor}, we combine these three shadow models' results ($n=3$), and construct user-level training samples to train a new combined auditor. This auditor is denoted as the \textit{Combined Auditor} that learns all kinds of popular ASR models. 

Fig.~\ref{fig:verify_train_alg} demonstrates the varied auditor performance (accuracy, precision and recall) when shadow models using various algorithms are deployed. For accuracy, all four auditors show an upward trend with a small training set size. The peak is observed at 500 training samples then decays to a stable smaller value at huge training set sizes. The GRU-based auditor surpasses the other three auditors in terms of accuracy, with the Combined Auditor performing the second-best when the auditor's training set size is smaller than 500. As for precision, all experiments show relatively high values (all above 60\%), particularly the LSTM-based auditor with a precision exceeding 80\%. According to Fig.~\ref{fig:verify_train_alg}c), the RNN-based auditor and GRU-based auditor show an upward trend in recalls. Herein, both of their recalls exceed 80\% when the training set size is larger than 500. The recall trends for the LSTM-based auditor and the Combined Auditor follows the opposite trend as that of GRU and RNN-based auditors. In general, the RNN-based auditor performs well across all three metrics. The LSTM-based auditor shows an excellent precision, while the GRU-based auditor obtains the highest accuracy. 

\begin{mdframed}[backgroundcolor=black!10,rightline=false,leftline=false,topline=false,bottomline=false,roundcorner=2mm] 
The algorithm selected for the shadow model will influence the auditor's performance. The Combined Auditor can achieve accuracy higher than the average, only if its training set is relatively small.
\end{mdframed}

\subsection{Effect of Noisy Queries}
\begin{figure*}[th]
  \centering
  \begin{subfigure}[b]{0.32\linewidth}
    \centering\includegraphics[width=\linewidth]{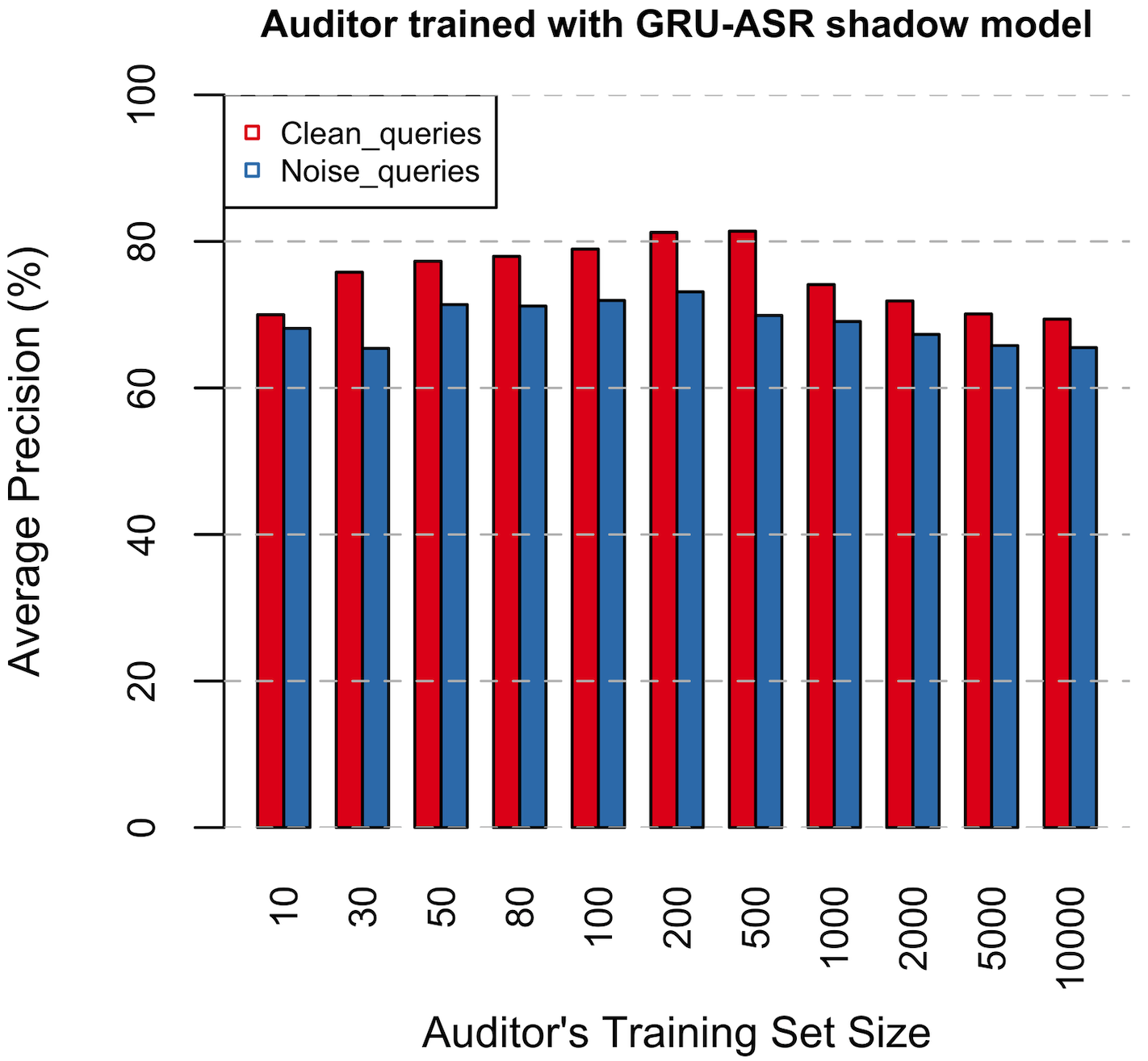}
    \caption{\centering GRU-based Auditor Precision}
  \end{subfigure}\hfill%
  \begin{subfigure}[b]{0.32\linewidth}
    \centering\includegraphics[width=\linewidth]{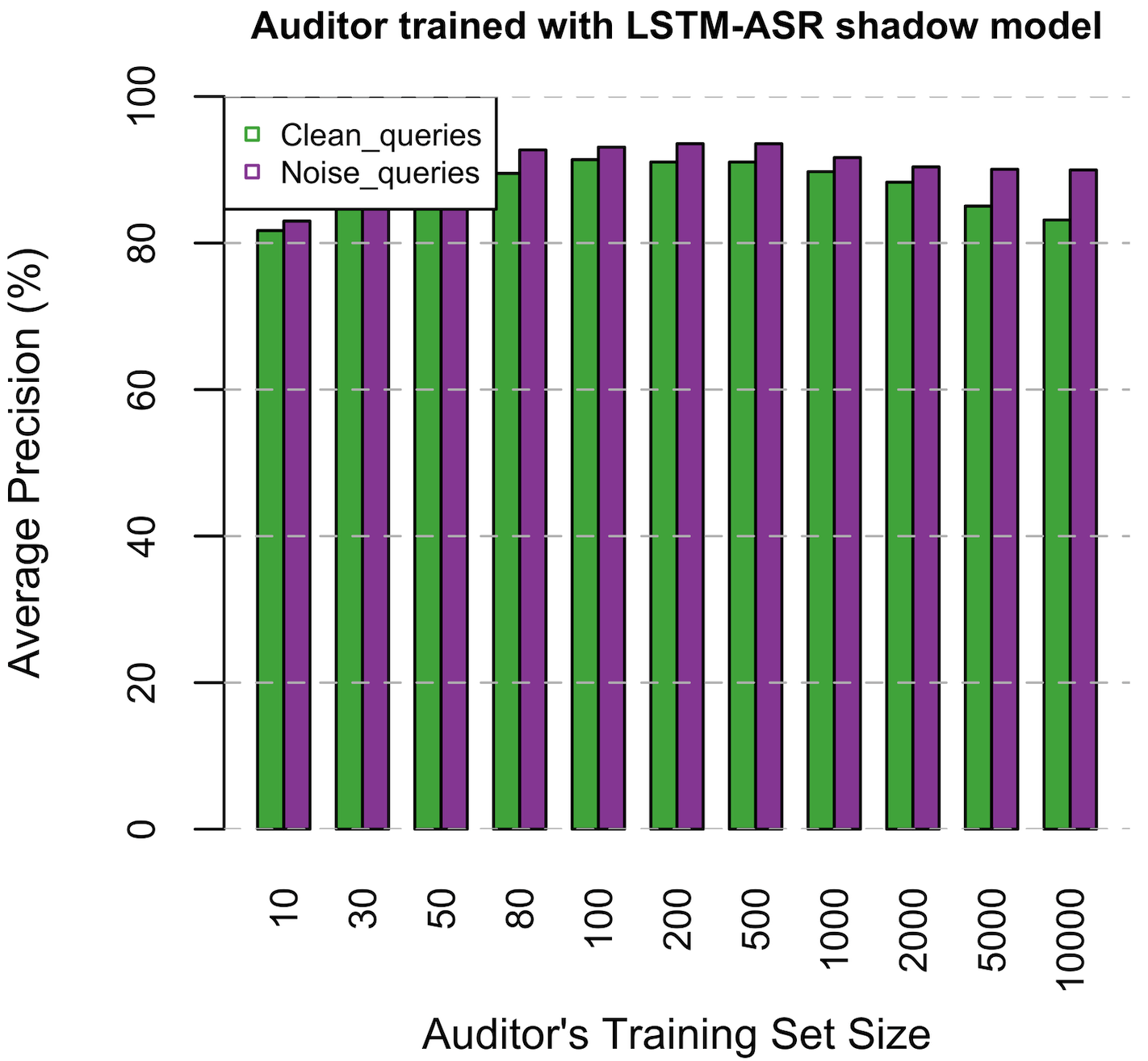}
    \caption{\centering LSTM-based Auditor Precision}
  \end{subfigure}%
  \begin{subfigure}[b]{0.32\linewidth}
    \centering\includegraphics[width=\linewidth]{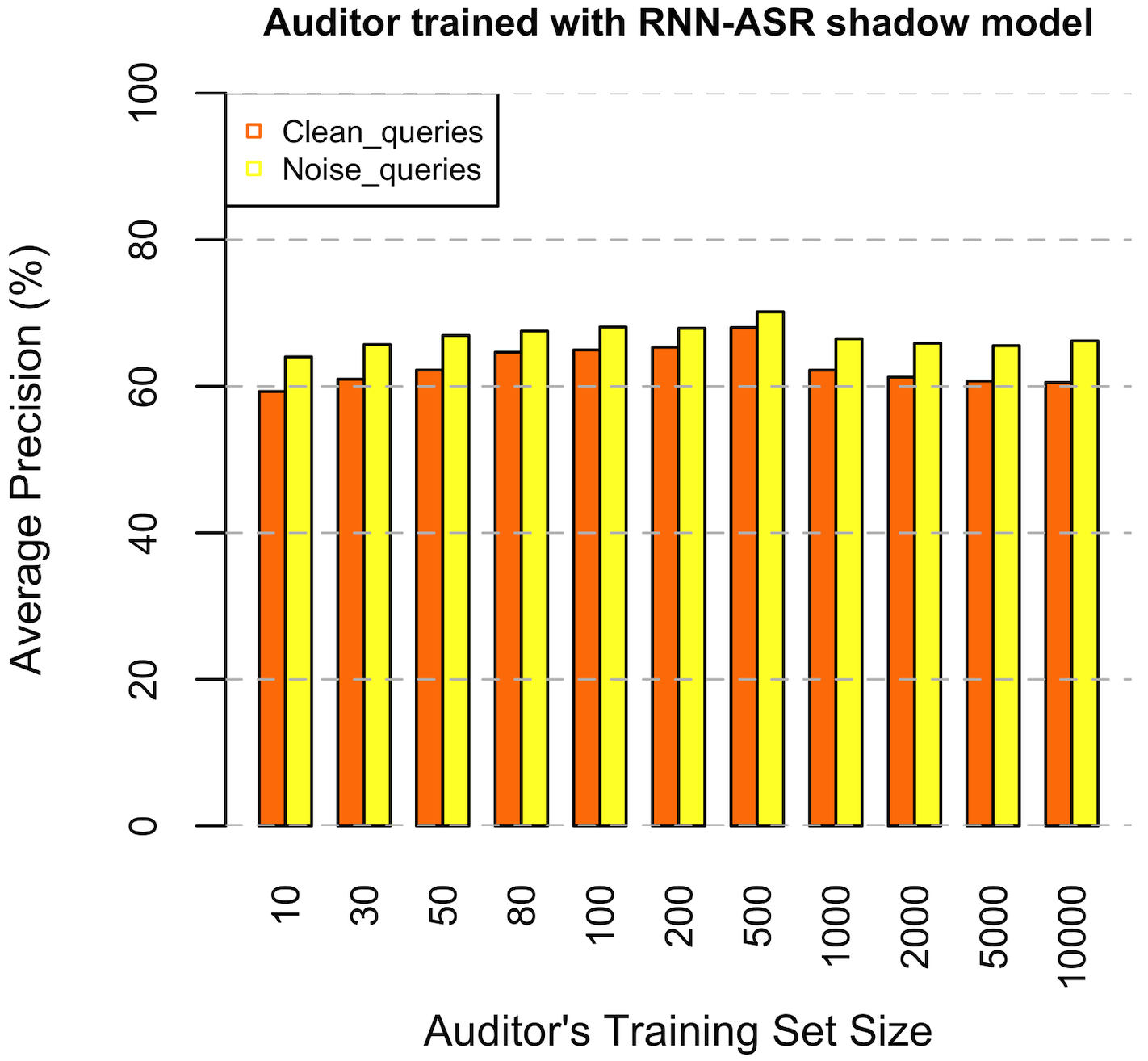}
    \caption{\centering RNN-based Auditor Precision}
  \end{subfigure}\hfill%
  
  \begin{subfigure}[b]{0.32\linewidth}
    \centering\includegraphics[width=\linewidth]{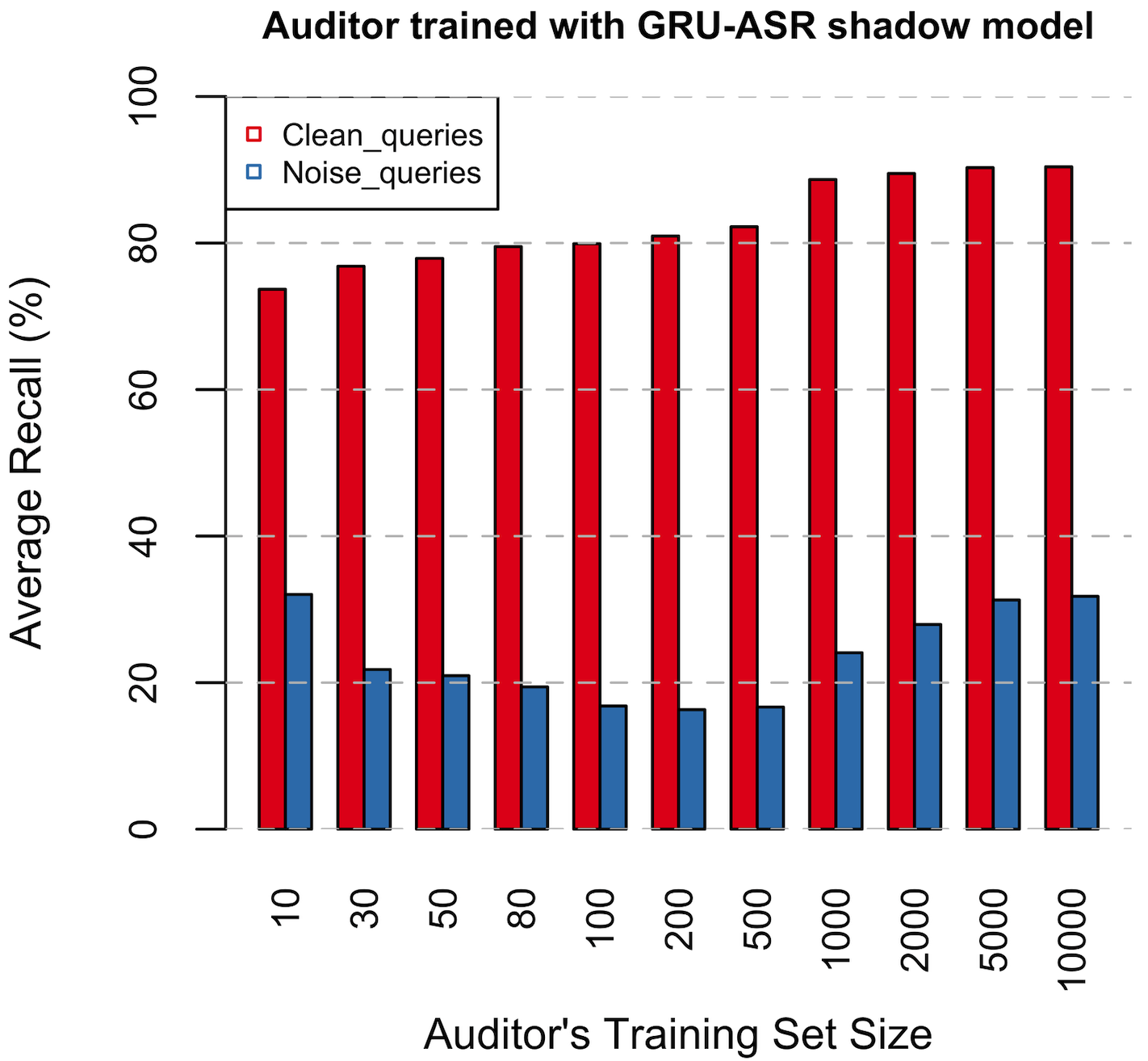}
    \caption{\centering GRU-based Auditor Recall}
  \end{subfigure}\hfill%
  \begin{subfigure}[b]{0.32\linewidth}
    \centering\includegraphics[width=\linewidth]{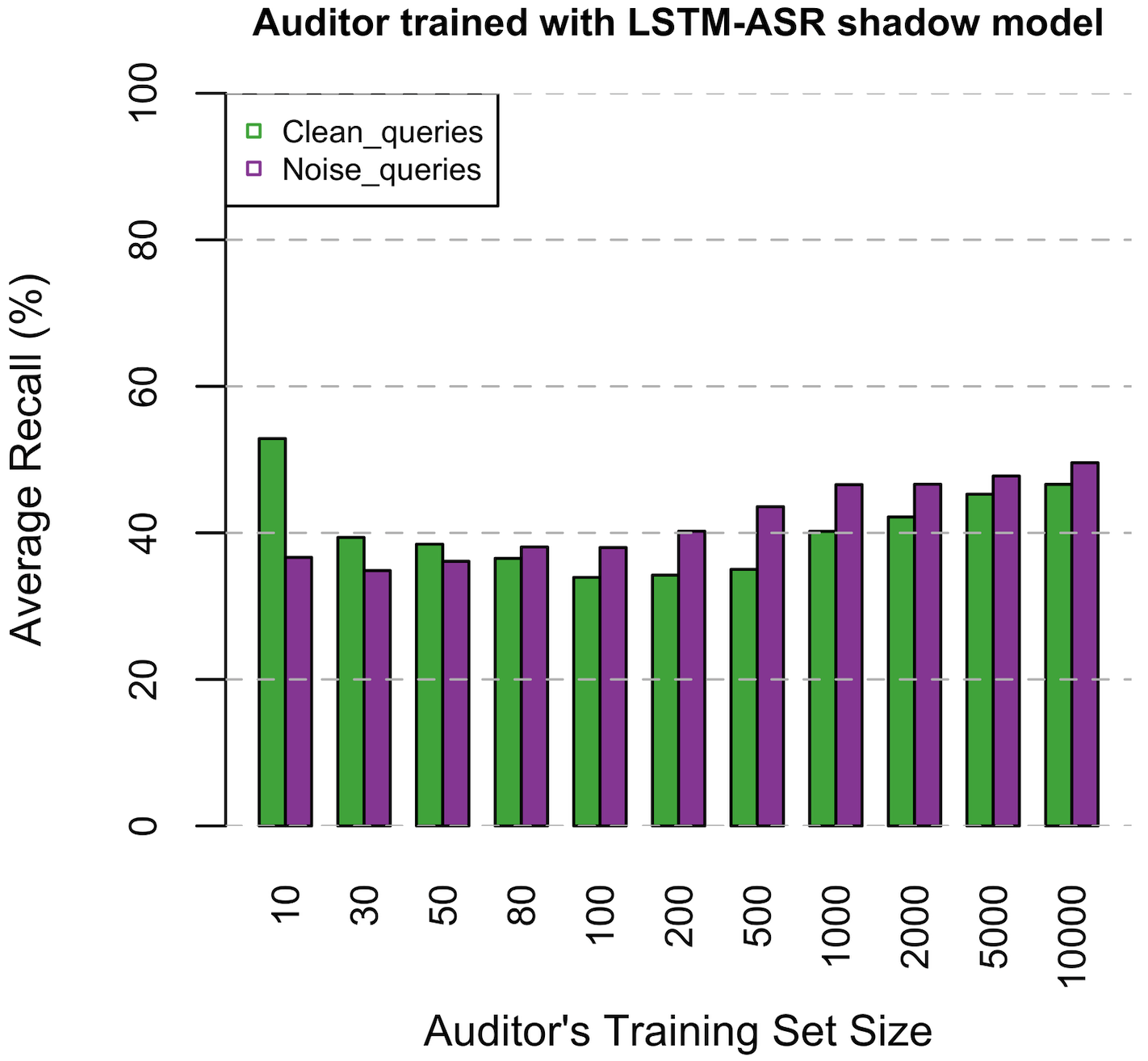}
    \caption{\centering LTSM-based Auditor Recall}
  \end{subfigure}\hfill%
  \begin{subfigure}[b]{0.32\linewidth}
    \centering\includegraphics[width=\linewidth]{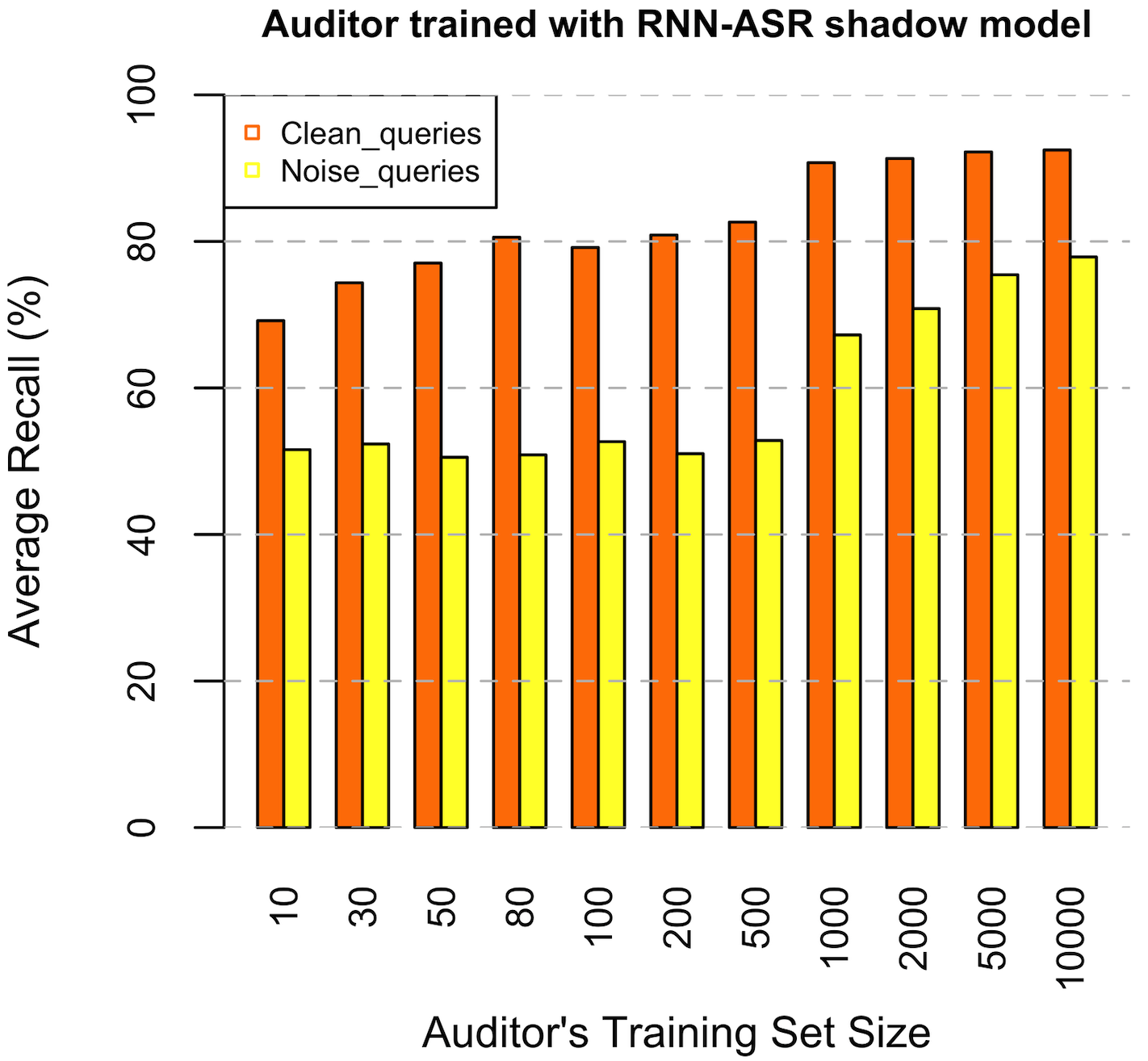}
    \caption{\centering RNN-based Auditor Recall}
  \end{subfigure}
  \caption{Different auditor audits noisy queries with different ASR shadow model.}
  \label{fig:verify_target_noisy}
\end{figure*}

An evaluation of the user-level audio auditor's robustness, when provided with noisy audios, is also conducted. We also consider the effect of the noisy audios when querying different kinds of auditors trained on different shadow model architectures. We shall describe the performance of the auditor with two metrics --- precision and recall; these results are illustrated in Fig.~\ref{fig:verify_target_noisy}.

To explore the effect of noisy queries, we assume that our target model is trained with noisy audios. Under the strict black-box access to this target model, we shall use different neural network structures to build the target model(s) and the shadow model(s). That is, the target model is an LSTM-ASR target model, while the GRU-ASR shadow model is used to train the GRU-based auditor. For evaluating the effect of the noisy queries, two target models are prepared using (i) clean audios (100 hours) and (ii) noisy audios (500 hours) as training sets. In addition to the GRU-based auditor, another two auditors are constructed, an LSTM-based auditor and an RNN-based auditor. The target models audited by the latter two auditors are the same as the GRU-based auditor. Herein, the LSTM-based auditor has an LSTM-ASR shadow model whose acoustic model shares the same algorithm as its LSTM-ASR target model.

Fig.~\ref{fig:verify_target_noisy}a and Fig.~\ref{fig:verify_target_noisy}d compare the precision and recall of the GRU-based auditor on target models trained with clean and noisy queries, respectively. Overall, the auditor's performance drops when auditing noisy queries, but the auditor still outperforms the random guess ($>$50\%). By varying the size of the auditor's training set, we observe the precision of the auditor querying clean and noisy audios displaying similar trends. When querying noisy audios, the largest change in precision is $\approx$11\%, where the auditor's training set size was 500. Its precision results of querying clean and noisy audios are around 81\% and 70\%, respectively. However, the trends of the two recall results are fairly the opposite, and noisy queries' recalls are decreasing remarkably. The lowest decent rate of the recall is about 42\%, where the auditor was trained with ten training samples. Its recall results of querying two kinds of audios are around 74\% and 32\%. In conclusion, we observe the impact of noisy queries on our auditor is fairly negative.

Fig.~\ref{fig:verify_target_noisy}b and Fig.~\ref{fig:verify_target_noisy}e display the LSTM-based auditor's precision and recall, respectively, while Fig.~\ref{fig:verify_target_noisy}c and Fig.~\ref{fig:verify_target_noisy}f illustrate the RNN-based auditor's performance. Similar trends are observed from the earlier precision results in the RNN-based auditor when querying clean and noisy queries. However, curiously, the RNN-based auditor, when querying noisy audios, slightly outperforms queries on clean audios. Similar to the noisy queries effect on GRU-based auditor, the noisy queries' recall of the RNN-based auditor decreases significantly versus the results of querying the clean audios. Though noisy queries show a negative effect, all recalls performed by the RNN-based auditor exceed 50\%, the random guess. As for the effect of noisy queries on the LSTM-based auditor, unlike GRU-based auditor and RNN-based auditor, the LSTM-based auditor demonstrates high robustness on noisy queries. For the most results of its precision and recall, the differences between the performance on clean and noisy queries are no more than 5\%.

\begin{mdframed}[backgroundcolor=black!10,rightline=false,leftline=false,topline=false,bottomline=false,roundcorner=2mm] 
In conclusion, noisy queries create a negative effect on our auditor's performance. Yet, if the shadow ASR model and the target ASR model are trained with the same algorithm, the negative effect can be largely eliminated.
\end{mdframed}

\subsection{Effect of Different ASR Model Pipelines on Auditor Performance} % V: architecture independent 

\begin{figure*}[th]
  \centering
  \begin{subfigure}[b]{0.32\linewidth}
    \centering\includegraphics[width=\linewidth]{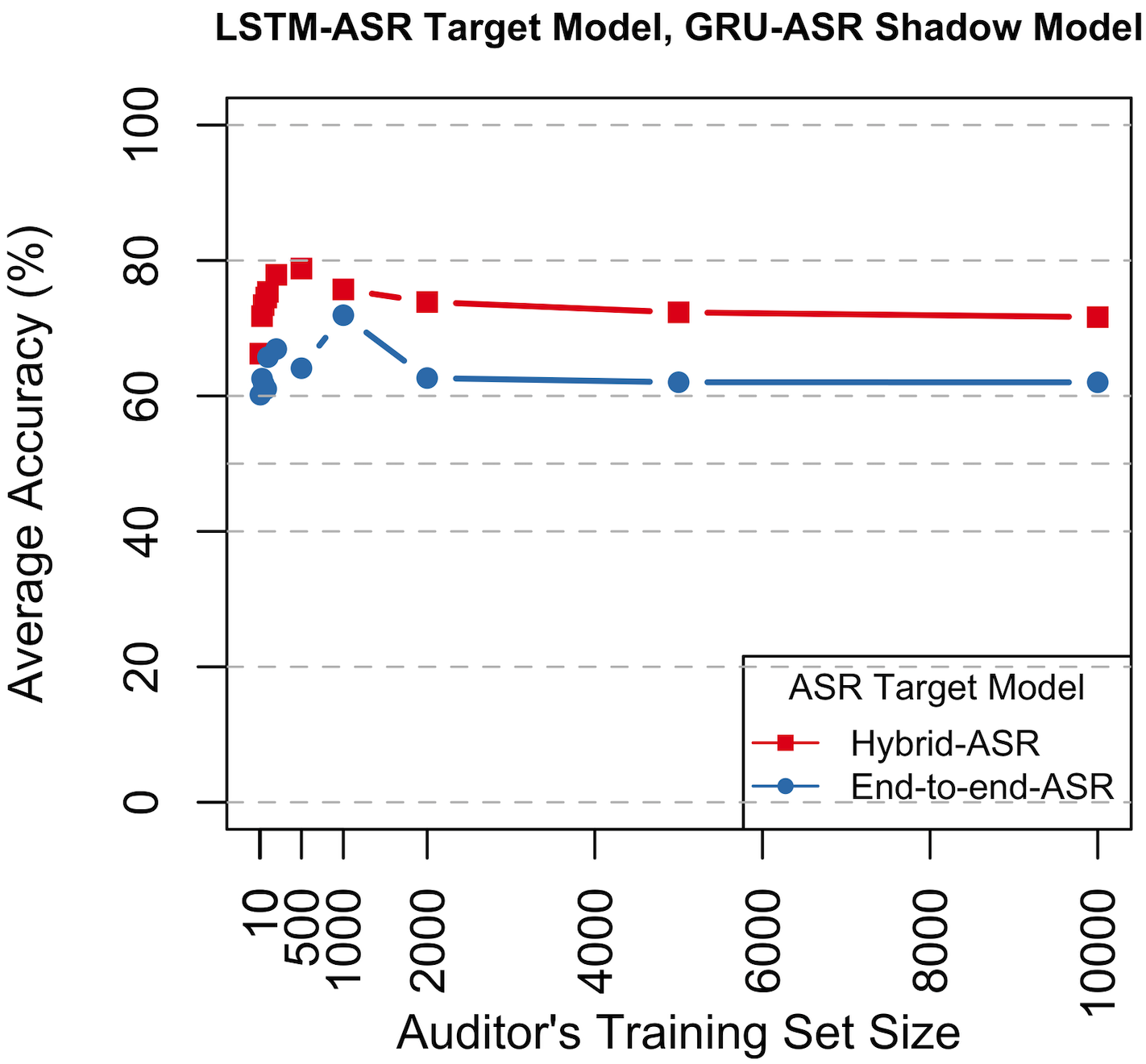}
    \caption{\centering Accuracy}
  \end{subfigure}\hfill%
  \begin{subfigure}[b]{0.32\linewidth}
    \centering\includegraphics[width=\linewidth]{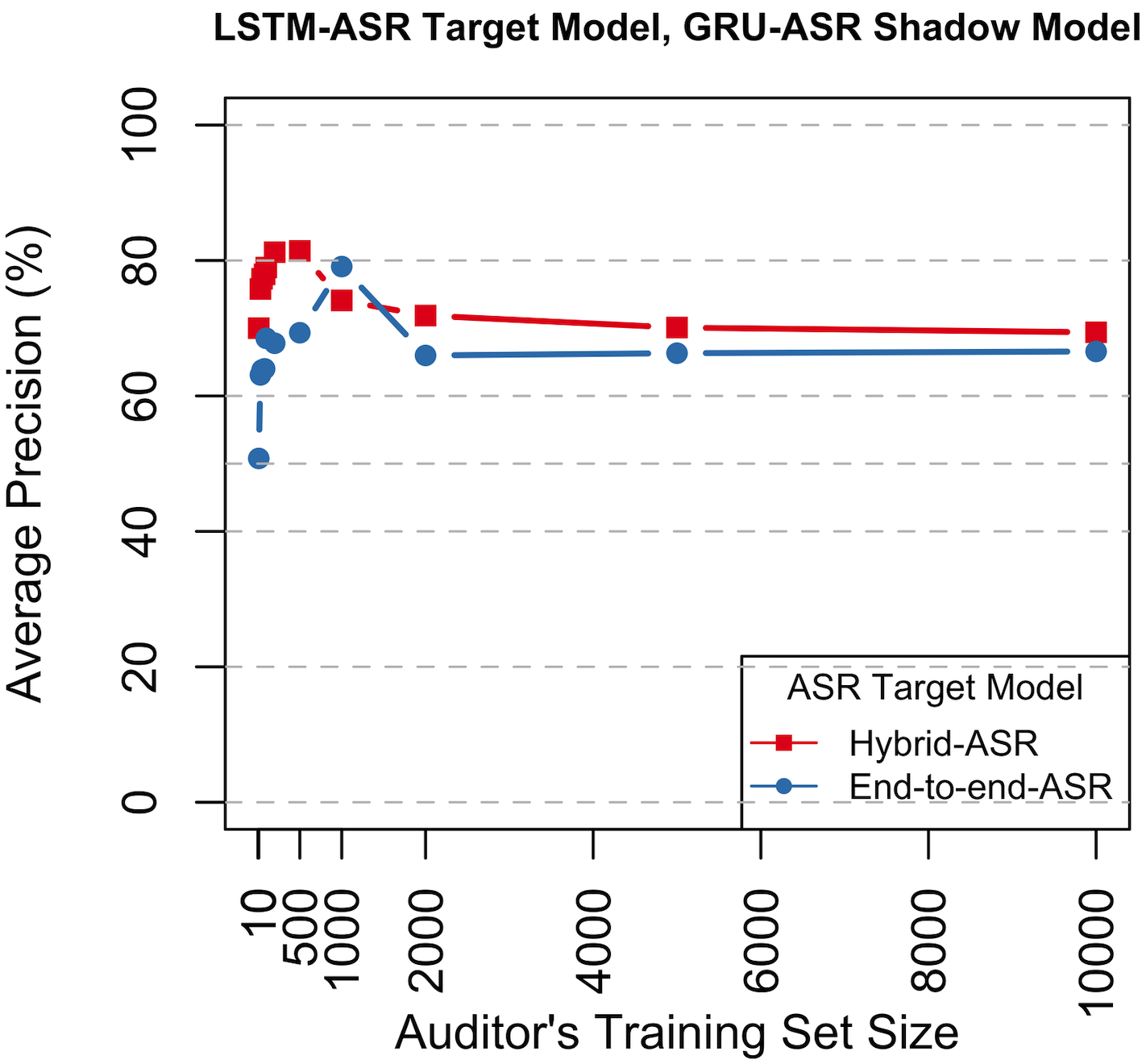}
    \caption{\centering Precision}
  \end{subfigure}\hfill%
  \begin{subfigure}[b]{0.32\linewidth}
    \centering\includegraphics[width=\linewidth]{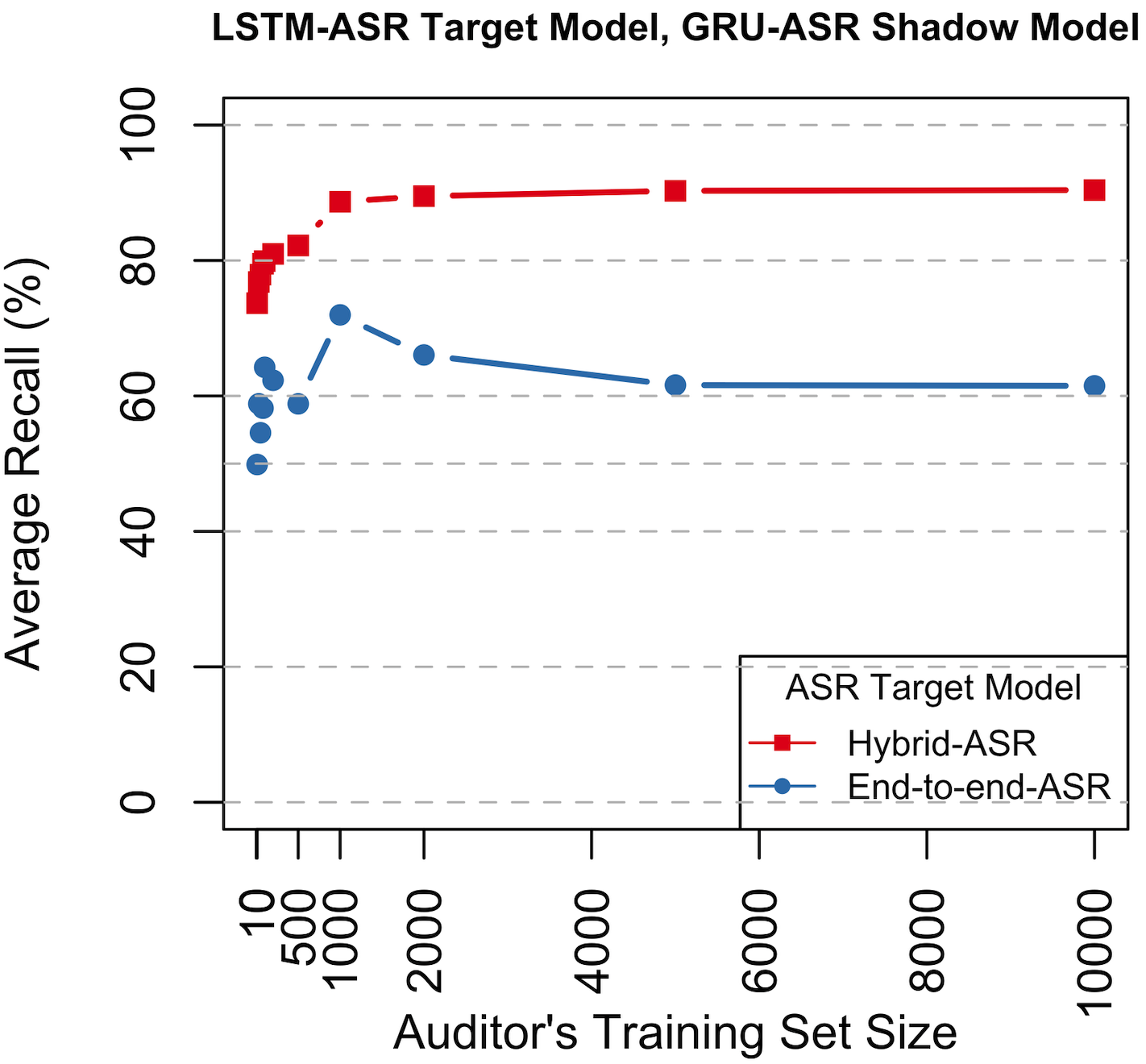}
    \caption{\centering Recall}
  \end{subfigure}
  \caption{The audit model audits different target ASR models trained with different pipelines.}
  \label{fig:verify_train_archi}
\end{figure*}

Aside from the ASR model's architecture, we examine the user-level auditor's robustness on different pipelines commonly found in ASR systems. In this section, the ASR pipeline is not just a machine learning, instead a complicated system, as shown in Fig.~\ref{fig:ASR_model}. In practice, the two most popular pipelines adopted in ASR systems are: a hybrid ASR system, or an end-to-end ASR system. We build our auditor using the GRU-ASR shadow model with two target models trained on systems built on the aforementioned ASR pipelines. Specifically, one target model utilizes the Pytorch-Kaldi toolkit to construct a hybrid DNN-HMM ASR system, while the other target model employs an end-to-end ASR system.

Fig.~\ref{fig:verify_train_archi} reports the performance (accuracy, precision, and recall) of the auditor when auditing the two different target pipelines. Overall, the auditor behaves well over all metrics when auditing either target models (all above 50\%). The auditor always demonstrates good performance when using a small number of training samples. The auditor targeting hybrid ASR in comparison to the end-to-end ASR target achieves a better result. A possible reason is that our auditor is constituted with a shadow model which has a hybrid ASR architecture. When focusing on the accuracy, the highest audit score of the hybrid ASR target model is 78.8\%, while that of the end-to-end ASR target model is 71.92\%. The difference in the auditor's precision is not substantially, with their highest precision scores as 81.4\% and 79.1\%, respectively. However, in terms of recall, the auditor's ability to determine the user-level membership on the hybrid ASR target model is much higher than the end-to-end target model, with maximum recall of 90\% and 72\%, respectively. 

When auditing the hybrid ASR target model we observed the model significantly outperforming other models. The training and testing data for both state-of-the-art ASR model architectures (i.e., hybrid and end-to-end) are the same. Thus, to confidently understand the impact of different ASR model pipelines on the auditor's performance, we shall also investigate the difference between the overfitting level of the hybrid ASR target model and that of the end-to-end ASR target model, as the overfitting of the model increases the success rate of membership inference attacks~\cite{shokri2017membership}.  Recall that overfitting was previously defined in Section~\ref{sec:exp}. The overfitting value of the hybrid ASR target model is measured as 0.04, while the overfitting level of the end-to-end ASR target model is 0.14. Contrary to the conclusions observed by \cite{salem2019ml}, the target model that was more overfit did not increase the performance of our user-level audio auditor. One likely reason is that our auditor audits the target model by considering user-level information under strict black-box access. Compared to conventional black-box access in \cite{salem2019ml}, our strict black-box access obtains its output from the transcribed text alone; consequently the influence of overfitting on specific words (WER) would be minimized. Thus, we can observe that our auditor's success is not entirely attributed to the degree of the target ASR model's overfitting alone.

\begin{mdframed}[backgroundcolor=black!10,rightline=false,leftline=false,topline=false,bottomline=false,roundcorner=2mm] 
In conclusion, different ASR pipelines between the target model and the shadow model negatively impact the performance of the auditor. Nevertheless, our auditor still performs well when the target model is trained following a different pipeline (i.e.,~an end-to-end ASR system),  significantly outperforming random guesses (50\%).
\end{mdframed}

\subsection{Real-World Audit Test} \label{subsec:real-world}
To test the practicality of our model in the real world, we keep our auditor model locally and conduct a proof-of-concept trial to audit iPhone Siri's speech-to-text service. We select the auditor trained by the GRU shadow model with LibriSpeech 360-hour voice data as its training set. To simplify the experiments, we sample five audios per user for each user's audit. According to the results presented in Fig.~\ref{fig:verify_queryN_A5all} and Fig.~\ref{fig:verify_queryN_all}, we select the auditor trained with five audios per user, where 1,000 users were sampled randomly as the auditor's training set. To gain the average performance of our auditor in the real world, we stored 100 auditors under the same settings with the training set constructed 100 times. The final performance is the average of these 100 auditors' results.

\noindent \textbf{Testbed and Data Preprocess.} The iPhone Siri provides strict black-box access to users, and the only dictation result is its predicted text. All dictation tasks were completed and all audios were recorded in a quiet surrounding. The clean audios were played via a Bluetooth speaker to ensure Siri can sense the audios. User-level features were extracted as per Section~\ref{subsec:overview}. The Siri is targeted on iPhone X of iOS 13.4.1. 

\noindent \textbf{Ground Truth.} We target a particular user $\hat{u}$ iPhone Siri's speech-to-text service. From Apple's privacy policy of Siri (see Appendix~D), iPhone user's Siri's recordings can be selected to improve Siri and dictation service in the long-term (for up to two years). We do note that this is an opt-in service. Simply put, this user can be labeled as ``member''. As for the ``nonmember'' user, we randomly selected 52 speakers from LibriSpeech dataset which was collected before  2014~\cite{panayotov2015librispeech}. As stated by the iPhone Siri's privacy policy, users' data ``may be retained for up to two years''. Thus, audios sampled from LibriSpeech can be considered out of this Siri's training set. We further regard the corresponding speakers of LibriSpeech as ``nonmember'' users. To avoid nonmember audios entering Siri's training data to retrain its ASR model during the testing time, each user's querying audios were completed on the same day we commenced tests for that user, with the Improve Siri \& Dictation setting turned off.

As we defined above, a ``member'' may make the following queries to our auditor: querying the auditor (i) with audios within the target model's training set ($D_{\hat{u}} = \bigcup {A}_{mem}^{in}$); (ii) with audios out of the target model's training set ($D_{\hat{u}} = \bigcup {A}_{mem}^{out}$); (iii) with part of his or her audios within the target model's training set ($D_{\hat{u}} = (\bigcup {A}_{mem}^{in}) \bigcup (\bigcup {A}_{mem}^{out})$). Thus, we generate six ``member'' samples where the audios were all recorded by the target iPhone's owner, including $D_{\hat{u}} = \bigcup \limits_{k=5} {A}_{mem}^{in}$, $D_{\hat{u}} = \bigcup \limits_{m=5} {A}_{mem}^{out}$, and $D_{\hat{u}} = (\bigcup \limits_{k} {A}_{mem}^{in}) \bigcup (\bigcup \limits_{m} {A}_{mem}^{out})$, where $k=1, 2, 3$ and $k+m=5$. In total, we collected 58 user-level samples with 6 ``member'' and 52 ``nonmember'' samples.

\noindent \textbf{Results.} We load 100 auditors to test those samples and the averaged overall accuracy as 89.76\%. Specifically, the average precision of predicting the ``member'' samples is 58.45\%, while the average precision of predicting the ``nonmember'' samples is 92.61\%. The average ROC AUC result is 72.6\%, which indicates our auditor's separability in this experiment. Except for the different behaviors of Siri translating the audios from ``member'' and ``nonmember'' users, we suspect that another reason of high precision on ``nonmember'' is due to the LibriSpeech audios are out of Siri's dictation scope. As for the low precision rate on ``member'' samples, we single out the data $D_{\hat{u}} = \bigcup \limits_{k=5} {A}_{mem}^{in}$ for testing. Additionally, its average accuracy result can reach 100\%; thus, the auditor is much more capable in handling ${A}_{mem}^{in}$ than ${A}_{mem}^{out}$, corroborating our observation in Section~\ref{sec:number of users}. 

\begin{mdframed}[backgroundcolor=black!10,rightline=false,leftline=false,topline=false,bottomline=false,roundcorner=2mm] 
In conclusion, our auditor shows a generally satisfying performance for users auditing a real-world ASR system, Apple's Siri on iPhone.
\end{mdframed}

\section{Threats to Auditors' Validity}
\noindent \textbf{Voiceprints Anonymization.} 
In determining the user-level membership of audios in the ASR model, our auditor relies on the target model's different behaviors when presented with training and unseen samples. The auditor's quality depends on the diverse responses of the target model when translating audio from different users. The feature is named users' voiceprints. The voiceprint is measured in \cite{qian2019speech} based on a speaker recognition system's accuracy. Our auditor represents the user's voiceprint according to two accumulated features, including missing characters and extra characters. However, if an ASR system is built using voice anonymization, our user-level auditor's performance would degrade significantly. The speaker's voice is disguised in \cite{qian2019speech} by using robust voice conversation while ensuring the correctness of speech content recognition. Herein, the most popular technique of voice conversation is frequency warping \cite{sundermann2003vtln}. In addition, abundant information about speakers' identities is removed in \cite{srivastava2019privacy} by using adversarial training for the audio content feature. Fig.~\ref{tab:verify_features} shows that the average accuracy of the auditor dropped by approximately 20\% without using the two essential features. Hence, auditing user-level membership in a speech recognition model trained with anonymized voiceprints remains as a future avenue of research.

\noindent \textbf{Differentially Private Recognition Systems.} 
Differential privacy (DP) is one of the most popular methods to prevent ML models from leaking any training data information. The work~\cite{song2019auditing} protects the text generative model by applying user-level DP to its language model. This method contains a language model during the hybrid ASR system's training, on which the user-level DP can be applied to obscure the identity, at the sacrifice of transcription performance. The speaker and speech characterization process is protected in \cite{nautsch2019preserving} by inserting noise during the learning process. However, due to strict black-box access and the lack of output probability information, our auditor's performance remains unknown on auditing the ASR model with DP. The investigation of our auditor's performance to this user protection mechanism is open for future research. 

\noindent \textbf{Workarounds and Countermeasures.}
Although Salem et al.~\cite{salem2019ml} has shown neither the shadow model nor the attack model are required to perform membership inference, due to the constraints of strict black-box access, the shadow model and auditor model approach provide a promising means to perform a more difficult task of user-level membership inference. Instead of the output probabilities, we mainly leverage the ASR model's translation errors in the character level to represent the model's behaviors. Alternative countermeasures against the membership inference, such as dropout, generally change the target model's output probability. However, the changes to probabilities of the ASR model's output are not as sensitive as changes to its translated text~\cite{salem2019ml}. Studying the extent of this sensitivity of ASR models remains as our future work. 

\noindent \textbf{Synthetic ASR Models.}
Another limitation of our work is that we evaluate our auditor on synthetic ASR systems trained on real-world datasets, and we have not applied the auditor to  extensive set of real-world models aside from Siri. However, we believe that our reconstruction of the ASR models closely mirrors ASR models in the wild.

\section{Related Work}\label{sec:lit}

\noindent \textbf{Membership Inference Attacks.}
% As a fundamental privacy threat to ML models, the membership inference
The attack distinguishes whether a particular data sample is a member of the target model's training set or not. Traditional membership inference attacks against ML models under black-box access leverage numerous shadow models to mimic the target model's behavior~\cite{shokri2017membership,long2018understanding,hayes2019logan}. Salem et al.~\cite{salem2019ml} revealed that membership inference attacks could be launched by directly utilizing the prediction probabilities and thresholds of the target model. Both works~\cite{long2018understanding} and \cite{yeom2018privacy} prove that overfitting of a model is sufficient but not a necessity to the success of a membership inference attack. Yeom et al.~\cite{yeom2018privacy} as well as Farokhi and Kaafar~\cite{farokhi2020modelling} formalize the membership inference attack with black-box and white-box access. All previously mentioned works consider record-level inference; however, Song and Shmatikov~\cite{song2019auditing} deploy a user-level membership inference attack in text generative models, with only the top-$n$ predictions known.

\noindent \textbf{Trustworthiness of ASR Systems.}
The ASR systems are often deployed on voice-controlled devices~\cite{chen2020devil}, voice personal assistants~\cite{shezanread}, and machine translation services~\cite{du2019sirenattack}. Tung and Shin~\cite{tung2019exploiting} propose \emph{SafeChat} to utilize a masking sound to distinguish authorized audios from unauthorized recording to protect any information leakage. Recent works~\cite{malik2019securing} and \cite{tom2018end} propose an audio cloning attack and audio reply attack against the speech recognition system to impersonate a legitimate user or inject unintended voices. Voice masquerading to impersonate users on the voice personal assistants has been studied~\cite{zhang2019dangerous}. Whereas Zhang et al.~\cite{zhang2019dangerous} propose another attack, namely \emph{voice squatting}, to hijack the user's voice command, producing a sentence similar to the legal command. Du et al.~\cite{du2019sirenattack} generate the adversarial audio samples to deceive the end-to-end ASR systems.

\noindent \textbf{Auditing ML Models.}
Many of the current proposed auditing services also seek to audit the bias and fairness of a given model~\cite{saleiro2018aequitas}. Works have also been presented to audit the ML model to learn and check the model's prediction reliability~\cite{schulam2019can, koh2017understanding, adler2018auditing}. Moreover, the auditor is utilized to evaluate the ML model's privacy risk when protecting an individual's digital rights~\cite{song2019auditing,miao2019audio}.

\noindent \textbf{Our Work.} Our user-level audio auditor audits the ASR model under the strict black-box access. As shown in Section~\ref{sec:audit}, we utilize the ASR model's translation errors in the character level to represent the model's behavior. Compared to related works under black-box access, our auditor does not rely on the target model's output probability~\cite{shokri2017membership,salem2019ml}. In addition, we sidestep the feature pattern of several top-ranked outputs of the target model adopted by Song and Shmatikov~\cite{song2019auditing}, instead we use one text output, the user's speed, and the input audio's true transcription, as we do not have access to the output probability (usually unattainable in ASR systems). Hence, our constraints of strict black-box access only allow accessing one top-ranked output. In this case, our user-level auditor (78.81\%) outperforms Song's and Shmatikov's user-level auditor (72.3\%) in terms of accuracy. Moreover, Hayes et al.~\cite{hayes2019logan} use adversarial generative networks (GANs) to approximate the target model's output probabilities while suffering big performance penalties with only 20\% accuracy, our auditor's accuracy is far higher. Furthermore, our auditor is much easier to be trained than the solution of finding outlier records with a unique influence on the target model \cite{long2018understanding}, because we only need to train one shadow model instead of many shadow (or reference) models.

% \section{Limitations and Future Work}
% \noindent \textbf{Further Investigation on Features.} From our set of selected features both audio-specific features and features capturing model behaviors performs well, as observed in our results. It remains to be seen if additional audio-specific features would specifically aid the task of user-level auditing. As there is a plethora of potential feature candidates, we consider this as part of future work.

% \noindent \textbf{Auditing Performance with Varied Numbers of Queries.} In our auditor, we observe that only a limited number of queries per user is necessary to audit the target ASR model, especially when the auditor is trained with a limited audios per user. 
% %during the auditing phase
% An interesting observation was that our user-level auditor's recall performance on more queries per user declines under the strict black-box access. We are continuing our investigation into why our auditor's ability to find unauthorized use of user data varies in this manner when being queried with different numbers of audios.

% \noindent \textbf{Member Audio in Siri Auditing.} In our setting, we make our best effort to ensure member audios are used for training. However, in our real-world evaluation, even with the ``Improve Siri \& Dictation'' setting turned on, with an extended period of continual use by our user, we cannot guarantee that member audios of our member user where actually used for training although we are confident they have been included. 

\section{Conclusion}
\label{sec:con}
This work highlights and exposes the potential of carrying out user-level membership inference audit in IoT voice services. The auditor developed in this paper has demonstrated promising data transferability, while allowing a user to audit his or her membership with a query of only nine audios. Even with audios are not within the target model's training set, the user's membership can still be faithfully determined. While our work has yet to overhaul the audit accuracy on various IoT applications across multiple learning models in the wild, we do narrow the gap towards defining clear membership privacy in the user level, rather than the record level~\cite{shokri2017membership}. However, questions remain about whether the privacy leakage hails from the data distribution or its intrinsic uniqueness of the record. More detail about limitations are stated in Appendices. Nevertheless, as we have shown, both a small training set size and the Combined Auditor, which combines results from various ASR shadow models to train the auditor, have a positive effect on the IoT audit model; on the contrary, audios recorded in a noisy environment and different ASR pipelines impose a negative effect on the given auditor; fortunately, the auditor still outperforms random guesses (50\%). Examining other performance factors on more real-world ASR systems in addition to our iPhone Siri trial and extending possible countermeasures against auditing are all worth further exploration. 

\section*{Acknowledgments}
We thank all anonymous reviewers for their valuable feedback. This research was supported by Australian Research Council, Grant No.~LP170100924. This work was also supported by resources provided by the Pawsey Supercomputing Centre, funded from the Australian Government and the Government of Western Australia.

% \presec
% \section*{Acknowledgement}
% \postsec

\bibliographystyle{abbrvnat}
\bibliography{ref_paper}

\section*{Appendices}

\subsection*{A. Datasets} \label{subsec:dataset}
The \noindent\textbf{LibriSpeech} speech corpus (LibriSpeech) contains 1,000 hours of speech audios from audiobooks which are part of the LibriVox project~\cite{panayotov2015librispeech}. This corpus is famous in training and evaluating speech recognition systems. At least 1,500 speakers have contributed their voices to this corpus. We use 100 hours of clean speech data with 29,877 recordings to train and test our target model. 360 hours of clean speech data, including 105,293 recordings, are used for training and testing the shadow models. Additionally, there are 500 hours of noisy data used to train the ASR model and to test our auditor's performance in a noisy environment.

The \noindent\textbf{TIMIT} speech corpus (TIMIT) is another famous speech corpus used to build ASR systems. This corpus recorded audios from 630 speakers across the United States, totaling 6,300 sentences \cite{garofolo1993timit}. In this work, we use all this data to train and test a target ASR model, and then audit this model with our auditor.

The \noindent\textbf{TED-LIUM} speech corpus (TED) collected audios based on TED Talks for ASR development \cite{rousseau2012ted}. This corpus was built from the TED talks of the IWSLT 2011 Evaluation Campaign. There are 118 hours of speeches with corresponding transcripts.

\subsection*{B. Evaluation Metrics} 
The user-level audio auditor is evaluated with four metrics calculated from the confusion matrix, which reports the number of true positives, true negatives, false positives and false negatives: \textbf{True Positive (TP)}, the number of records we predicted as ``\texttt{member}'' are correctly labeled; \textbf{True Negative (TN)}, the number of records we predicted as ``\texttt{nonmember}'' are correctly labeled; \textbf{False Positive (FP)}, the number of records we predicted as ``\texttt{member}'' are incorrectly labeled; \textbf{False Negative (FN)}, the number of records we predicted as ``\texttt{nonmember}'' are incorrectly labeled. Our evaluation metrics are derived from the above-mentioned numbers.
\begin{itemize}
\item Accuracy: the percentage of records correctly classified by the auditor model. 
\item Precision: the percentage of records correctly determined as ``\texttt{member}'' by the auditor model among all records determined as ``\texttt{member}''. 
\item Recall: the percentage of all true ``\texttt{member}'' records correctly determined as ``\texttt{member}''.
\item F1-score: the harmonic mean of precision and recall. 
\end{itemize} 

\subsection*{C. ASR Models' Architectures}
On the LibriSpeech 360-hour voice dataset, we build one GRU-ASR model with the Pytorch-Kaldi toolkit. That is, we train a five-layer GRU network with each hidden layer of size 550 and one Softmax layer. We use \texttt{tanh} as the activation function. The optimization function is Root Mean Square Propagation (RMSProp). We set the learning rate as 0.0004, the dropout rate for each GRU layer 0.2, and the number of epochs of training 24.

On the LibriSpeech 360-hour voice dataset, we train another ASR model using the Pytorch-Kaldi toolkit. Specifically, it is a four-layer RNN network with each hidden layer of size 550 using ReLU as the activation function and a Softmax layer. The optimization function is RMSProp. We set the learning rate  as 0.00032, the dropout rate for each RNN layer 0.2, and the number epochs of training 24.

On the LibriSpeech 360-hour voice dataset, we train one hybrid LSTM-ASR model. The acoustic model is constructed with a four-layer LSTM and one Softmax layer. The size of each hidden LSTM layer is 550 along with 0.2 dropout rate. The activation function is \texttt{tanh}, while the optimization function is RMSProp. The learning rate is 0.0014, and the maximum number of training epochs is 24.

On the LibriSpeecch 100-hour voice dataset, we train a hybrid ASR model. The acoustic model is constructed with a four-layer LSTM and one Softmax layer. Each hidden LSTM layer has 550 neurons along with 0.2 dropout rate. The activation function is \texttt{tanh}, while the optimization function is RMSProp. The learning rate is 0.0016, and the maximum number of training epochs is 24.

On the LibriSpeecch 100-hour voice dataset, we train an end-to-end ASR model. The encoder is constructed with a five-layer LSTM with each layer of size 320 and with 0.1 dropout rate. We use one layer location-based attention with 300 cells. The decoder is constructed with a one-layer LSTM with 320 neurons along with 0.5 dropout rate. The CTC decoding is enabled with a weight of 0.5. The optimization function is Adam. The learning rate is 1.0, and the total number of training epochs is 24.

On the TEDLium dataset, we train a hybrid ASR model. The acoustic model is constructed with a four-layer LSTM and one Softmax layer. Each hidden LSTM layer has 550 neurons along with 0.2 dropout rate. The activation function is \texttt{tanh}, while the optimization function is RMSProp. The learning rate is 0.0016, and the number of maximum training epochs is 24.

On the TIMIT dataset, we train a hybrid ASR model. The acoustic model is constructed with a four-layer LSTM and one Softmax layer. Each hidden LSTM layer has 550 neurons along with 0.2 dropout rate. The activation function is \texttt{tanh}, while the optimization function is RMSProp. The learning rate is 0.0016, and the number of maximum training epochs is 24.

\subsection*{D. Real-World Audit Test}\label{apdx:D}

\begin{figure}[t]
\centering
\includegraphics[width=0.65\columnwidth]{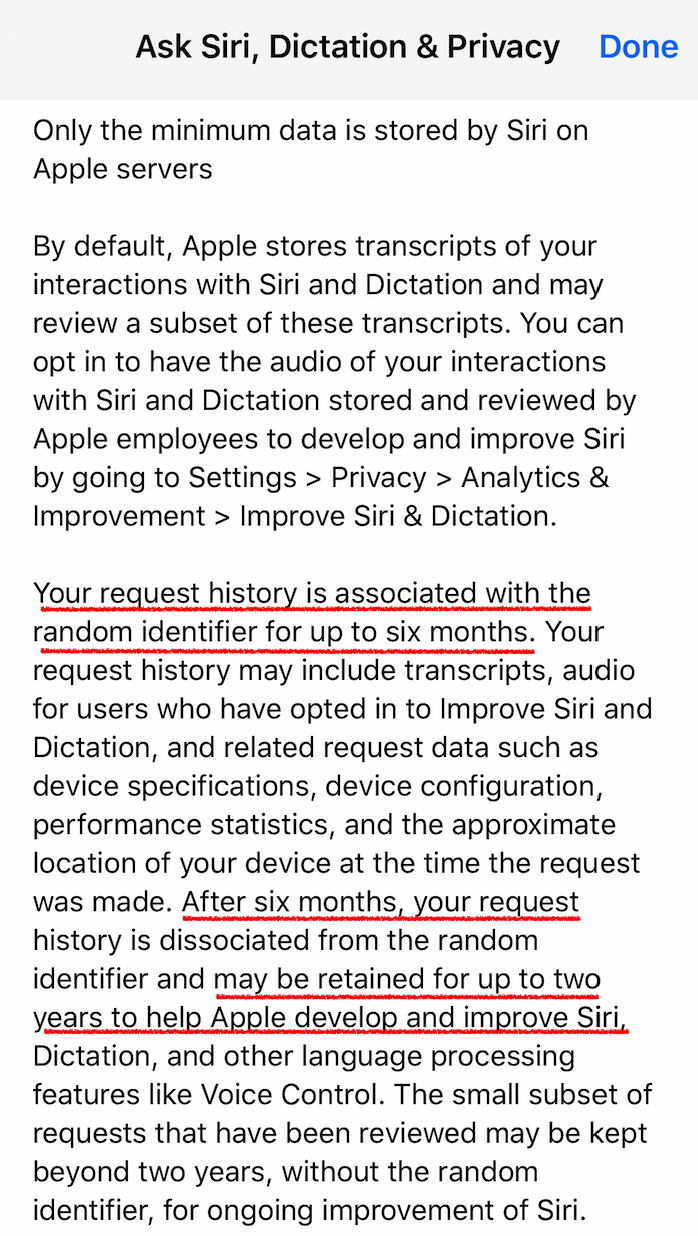}
\vspace{-2mm}
\caption{The privacy policy of Apple's Siri}
\label{fig:siri_privacy}
\end{figure}

Siri is a virtual assistant provided by Apple in their iOS, iPadOs, and macOS.  Siri's natural language interface is used to answer users' voice queries and make recommendations~\cite{siri2011}. The privacy policy of Apple's Siri is shown in Fig.~\ref{fig:siri_privacy}. The user, who is considered as a member user of Siri's ASR model in our setting, has used the targeted iPhone for more than two years, frequently interacting with Siri, often with the Improve Siri \& Dictation service opted in. As for the member user's member audio, we carefully chose five phrases that the user had certainly used when engaging with Siri. Starting with ``Hey Siri'' and phrases from common interactions including ``Hey Siri'', ``What's the weather today'', ``What date is it today'', ``Set alarm at 10 o'clock'', and ``Hey Siri, what's your name''. As for the member user's non-member audio, we chose five short phrases in LibriSpeech that the user had never used to interact with Siri (e.g. ``we ate at many men's tables uninvited''). These phrases were recorded using the member user's voice along with our member user's non-member audios. The use of either set of phrases should produce an ability to audit as recall that our method imitates the user as a whole when auditing the model, irrespective of whether a specific audio phrase was used to train/update this model. Lastly, for the nonmember users' nonmember audio, the target Siri's language is English (Australia). Since the LibriSpeech dataset was collected before 2014 and as stated by the iPhone Siri's privacy policy, users' data ``may be retained for up to two years''. We consider these recordings to not be part of the Siri training dataset, and thus we have nonmember users' nonmember audio. We further assume that our selected user phrases about book reading are nonmembers.

\subsection*{E. Limitations and Future Work}
\noindent \textbf{Further Investigation on Features.} From our set of selected features both audio-specific features and features capturing model behaviors performs well, as observed in our results. It remains to be seen if additional audio-specific features would specifically aid the task of user-level auditing. As there are many potential feature candidates, we consider this as part of future work.

\noindent \textbf{Auditing Performance with Varied Numbers of Queries.} In our auditor, we observe that only a limited number of queries per user is necessary to audit the target ASR model, especially when the auditor is trained with a limited audios per user. 
%during the auditing phase
An interesting observation was that our user-level auditor's recall performance on more queries per user declines under the strict black-box access. We are continuing our investigation into why our auditor's ability to find unauthorized use of user data varies in this manner when being queried with different numbers of audios.

\noindent \textbf{Member Audio in Siri Auditing.} In our setting, we make our best effort to ensure member audios are used for training. However, in our real-world evaluation, even with the ``Improve Siri \& Dictation'' setting turned on, with an extended period of continual use by our user, we cannot guarantee that member audios of our member user where actually used for training although we are confident they have been included. 

\end{document}